%% file: metmet_arXiv.tex
\documentclass[aps,jcp,superscriptaddress,amsmath,amssymb]{revtex4-1}

\usepackage{natbib}
\setcitestyle{super}

\usepackage{graphicx}
\usepackage{dcolumn}
\usepackage{bm}

\newcolumntype{.}{D{.}{.}{8}}

\usepackage[utf8]{inputenc}
\usepackage{physics}
\usepackage{amsmath, amssymb}
\usepackage{tabularx,booktabs,array,dcolumn}
\usepackage{setspace}
\usepackage{vmargin}
\usepackage{xcolor}
\usepackage{graphicx,psfrag,subfigure}

\usepackage{float}
\usepackage[all]{xy}
\usepackage{multirow}

\usepackage[round-mode=places,round-precision=4]{siunitx} %change the number of digits printed in the PDF
\newcommand{\mynum}[2]{\num[round-mode=places,round-precision=#1]{#2}}

\usepackage{bm}
\usepackage{mathtools}
\usepackage{physics}

\newcommand{\iim}{\mathrm{i}}

\newcommand{\cm}{cm$^{-1}$}

\usepackage[unicode]{hyperref}
\hypersetup{
   unicode=true,          % non-Latin characters in Acrobat??s bookmarks
   plainpages=false,
   colorlinks=true,       % false: boxed links; true: colored links
   citecolor=blue,        % color of links to bibliography
}

\def\metmet{(CH$_4$)$_2$}
\def\metwat{CH$_4\cdot$H$_2$O}

\def\xone{\text{X}_1(1)}
\def\xtwo{\text{X}_2(1)}
\def\xthree{\text{X}_3(1)}
\def\xfour{\text{X}_4(1)}
\def\xfive{\text{X}_5(2)}
\def\xsix{\text{X}_6(2)}
\def\xseven{\text{X}_7(2)}
\def\xeight{\text{X}_8(2)}
\def\xnine{\text{X}_9(2)}
\def\xten{\text{X}_{10}(6)}
\def\xeleven{\text{X}_{11}(6)}
\def\xtwelve{\text{X}_{12}(9)}
\def\xthirteen{\text{X}_{13}(9)}
\def\xfourteen{\text{X}_{14}(9)}
\def\xfifteen{\text{X}_{15}(9)}
\def\xsixteen{\text{X}_{16}(12)}
\usepackage{soul}
\usepackage{color}

\def\metmet{(CH$_4$)$_2$}
\def\metwat{CH$_4\cdot$H$_2$O}

\def\xone{\text{X}_1(1)}
\def\xtwo{\text{X}_2(1)}
\def\xthree{\text{X}_3(1)}
\def\xfour{\text{X}_4(1)}
\def\xfive{\text{X}_5(2)}
\def\xsix{\text{X}_6(2)}
\def\xseven{\text{X}_7(2)}
\def\xeight{\text{X}_8(2)}
\def\xnine{\text{X}_9(2)}
\def\xten{\text{X}_{10}(6)}
\def\xeleven{\text{X}_{11}(6)}
\def\xtwelve{\text{X}_{12}(9)}
\def\xthirteen{\text{X}_{13}(9)}
\def\xfourteen{\text{X}_{14}(9)}
\def\xfifteen{\text{X}_{15}(9)}
\def\xsixteen{\text{X}_{16}(12)}
\newcommand{\bos}[1]{\pmb{#1}}

\def\metmet{(CH$_4$)$_2$}
\def\metwat{CH$_4\cdot$H$_2$O}

\def\xone{\text{X}_1(1)}
\def\xtwo{\text{X}_2(1)}
\def\xthree{\text{X}_3(1)}
\def\xfour{\text{X}_4(1)}
\def\xfive{\text{X}_5(2)}
\def\xsix{\text{X}_6(2)}
\def\xseven{\text{X}_7(2)}
\def\xeight{\text{X}_8(2)}
\def\xnine{\text{X}_9(2)}
\def\xten{\text{X}_{10}(6)}
\def\xeleven{\text{X}_{11}(6)}
\def\xtwelve{\text{X}_{12}(9)}
\def\xthirteen{\text{X}_{13}(9)}
\def\xfourteen{\text{X}_{14}(9)}
\def\xfifteen{\text{X}_{15}(9)}
\def\xsixteen{\text{X}_{16}(12)}

\usepackage[round-mode=places,round-precision=4]{siunitx} %change the number of digits printed in the PDF
\definecolor{cream}{RGB}{222,217,201}

\usepackage[unicode]{hyperref}
\usepackage{soul}
\hypersetup{
   unicode=true,          % non-Latin characters in Acrobat??s bookmarks
   plainpages=false,
   colorlinks=true,       % false: boxed links; true: colored links
   linkcolor=blue,          % color of internal links
   citecolor=blue,        % color of links to bibliography
}

\usepackage{url}

\begin{document}

\title{%
Methane dimer rovibrational states and Raman transition moments
}

\author{Alberto Mart\'in Santa Dar\'ia}
\affiliation{%
Departamento de Química Física, University of Salamanca,
37008 Salamanca, Spain}
\affiliation{%
ELTE, E\"otv\"os Lor\'and University, 
Institute of Chemistry, 
P\'azm\'any P\'eter s\'et\'any 1/A,
1117 Budapest, Hungary}

\author{Gustavo Avila}
\affiliation{%
ELTE, E\"otv\"os Lor\'and University, 
Institute of Chemistry, 
P\'azm\'any P\'eter s\'et\'any 1/A,
1117 Budapest, Hungary}

\author{Edit M\'atyus}
\email{edit.matyus@ttk.elte.hu}
\affiliation{%
ELTE, E\"otv\"os Lor\'and University, 
Institute of Chemistry, 
P\'azm\'any P\'eter s\'et\'any 1/A,
1117 Budapest, Hungary}

\date{\today}
\begin{abstract}
\noindent 
Benchmark-quality rovibrational data are reported for the methane dimer from variational nuclear motion computations using
an \emph{ab initio} intermolecular potential energy surface reported by [M. P. Metz \textit{et al.}, \textit{Phys. Chem. Chem. Phys.}, 2019, \textbf{21}, 13504-13525].
A simple polarizability model is used to compute Raman transition moments that may be relevant for future direct observation of the intermolecular dynamics. Non-negligible $\Delta K\neq 0$ transition moments arise in this symmetric top system due to strong rovibrational couplings.
\end{abstract}

\maketitle 

\clearpage

%%%%%%%%%%%%%%%%%%%%%%%%%%%%%%%%%%%%%%%%%%%%%%%%%%%%%%%%%%%%%%%%%%%%%%%%%%%%%%%%%%%%%%%
\input{metmet_text}

\clearpage
%\bibliography{mybib}


%%%%%%%%%%%%%%%%%%%%%%%%%%%%%%%%%%%%%%%%%%%%%%%%%%%%%%%%%%%%%%%%%%%%%%%%%%%%%%%%%%%%%%%
\clearpage
\setcounter{section}{0}
\renewcommand{\thesection}{S\arabic{section}}
\setcounter{subsection}{0}
\renewcommand{\thesubsection}{S\arabic{section}.\arabic{subsection}}

\setcounter{equation}{0}
\renewcommand{\theequation}{S\arabic{equation}}

\setcounter{table}{0}
\renewcommand{\thetable}{S\arabic{table}}

\setcounter{figure}{0}
\renewcommand{\thefigure}{S\arabic{figure}}

~\\[0.cm]
\begin{center}
\begin{minipage}{1.\linewidth}
\centering
\textbf{Supplementary Material} \\[0.5cm]

\textbf{\large Methane dimer rovibrational states and Raman transition moments}
\end{minipage}
~\\[0.5cm]
\begin{minipage}{0.6\linewidth}
\centering

Alberto Mart\'in Santa Dar\'ia,$^{1,2}$ Gustavo Avila,$^2$ and \\ Edit M\'atyus$^{2,\ast}$ \\[0.15cm]

$^1$~\emph{Departamento de Qu\'imica F\'isica, Universidad de Salamanca, Spain} \\
$^2$~\emph{ELTE, Eötvös Loránd University, Institute of Chemistry, 
Pázmány Péter sétány 1/A, Budapest, H-1117, Hungary} \\[0.15cm]
$^\ast$ edit.matyus@ttk.elte.hu \\
\end{minipage}
~\\[0.15cm]
(Dated: March 8, 2024)
\end{center}

\clearpage
%\tableofcontents
%%%%%%%%%%%%%%%%%%%%%%%%%%%%%%%%%%%%%%%%%%%%%%%%%%%%%%%%%%%%%%%%%%%%%%%%%%%%%%%%%%%%%%%
\section{Molecular symmetry group of the methane dimer}

\begin{table*}[h!]
  \caption{%
    Character table of the molecular symmetry group of the methane dimer. 
    The table was generated using the GAP program according to instructions of 
    Ref.~\cite{ScLe04}.
    \label{tab:g576chartab}    
  }
  \centering
  \scalebox{0.94}{
  \begin{tabular}{@{}l@{\ \ }cc@{}c c@{}c c@{}c c@{}c c@{}c c@{}c c@{}c c@{}c c@{}c c@{}c c@{}c c@{}c c@{}c c@{}c c@{}c c@{\ \ }c@{}}
    \hline\\[-0.4cm]\hline\\[-0.4cm]
{$G_{576}^\text{a}$}  & $\hat{E}$ &
\multicolumn{2}{c}{\rotatebox{90}{(14)(23)}}	&
\multicolumn{2}{c}{\rotatebox{90}{(14)(23)(58)(67)}}	&
\multicolumn{2}{c}{\rotatebox{90}{(243)}}	&
\multicolumn{2}{c}{\rotatebox{90}{(243)(58)(67)}}	&
\multicolumn{2}{c}{\rotatebox{90}{(243)(687)}}	&
\multicolumn{2}{c}{\rotatebox{90}{(234)(687)}}	&
\multicolumn{2}{c}{\rotatebox{90}{(34)(78)(9,10)}}	&
\multicolumn{2}{c}{\rotatebox{90}{(1423)(78)(9,10)}}	&
\multicolumn{2}{c}{\rotatebox{90}{(1423)(5867)(9,10)}}	&
\multicolumn{2}{c}{\rotatebox{90}{(15)(26)(37)(48)}}	&
\multicolumn{2}{c}{\rotatebox{90}{(1548)(2637)}}	&
\multicolumn{2}{c}{\rotatebox{90}{(15)(264837)}}	&
\multicolumn{2}{c}{\rotatebox{90}{(15)(26)(38)(47)(9,10)}}	&
\multicolumn{2}{c}{\rotatebox{90}{(1547)(2638)(9,10)}}	&
\multicolumn{2}{c}{\rotatebox{90}{(15)(264738)(9,10)}}	\\
% \cline{1-32}\\[-0.4cm]
%
$R$$^\text{b}$   & 	1	&&	2	&&	3	&&	4	&&	5	&&	6	&&	7	&&	8	&&	9	&&	10	&&	11	&&	12	&&	13	&&	14	&&	15	&&	16	\\
% \cline{1-32}\\[-0.4cm]
%
$K_R$$^\text{c}$ & 	1	&&	6	&&	9	&&	16	&&	48	&&	32	&&	32	&&	36	&&	72	&&	36	&&	12	&&	36	&&	96	&&	12	&&	36	&&	96	&& $g_\mathrm{ns}$$^\text{d}$ \\
\cline{1-34}\\[-0.4cm]
$\xone$	         & 	1	&&	1	&&	1	&&	1	&&	1	&&	1	&&	1	&&	1	&&	1	&&	1	&&	1	&&	1	&&	1	&&	1	&&	1	&&	1	&& 15\\
$\xtwo$	         &  	1	&&	1	&&	1	&&	1	&&	1	&&	1	&&	1	&	$-$&1	&	$-$&1	&	$-$&1	&	$-$&1	&	$-$&1	&	$-$&1	&&	1	&&	1	&&	1 && 10	\\
$\xthree$	     & 	1	&&	1	&&	1	&&	1	&&	1	&&	1	&&	1	&	$-$&1	&	$-$&1	&	$-$&1	&&	1	&&	1	&&	1	&	$-$&1	&	$-$&1	&	$-$&1 && 15	\\
$\xfour$	     & 	1	&&	1	&&	1	&&	1	&&	1	&&	1	&&	1	&&	1	&&	1	&&	1	&	$-$&1	&	$-$&1	&	$-$&1	&	$-$&1	&	$-$&1	&	$-$&1 && 10	\\
$\xfive$	     & 	2	&&	2	&&	2	&	$-$&1	&	$-$&1	&&	2	&	$-$&1	&&	.	&&	.	&&	.	&&	.	&&	.	&&	.	&&	2	&&	2	&	$-$&1 && 1	\\
$\xsix$	         & 	2	&&	2	&&	2	&	$-$&1	&	$-$&1	&&	2	&	$-$&1	&&	.	&&	.	&&	.	&&	.	&&	.	&&	.	&	$-$&2	&	$-$&2	&&	1 && 1	\\
$\xseven$	     & 	2	&&	2	&&	2	&	$-$&1	&	$-$&1	&	$-$&1	&&	2	&&	.	&&	.	&&	.	&	$-$&2	&	$-$&2	&&	1	&&	.	&&	.	&&	. &&    0	\\
$\xeight$	     & 	2	&&	2	&&	2	&	$-$&1	&	$-$&1	&	$-$&1	&&	2	&&	.	&&	.	&&	.	&&	2	&&	2	&	$-$&1	&&	.	&&	.	&&	. && 2	\\
$\xnine$	     & 	4	&&	4	&&	4	&&	1	&&	1	&	$-$&2	&	$-$&2	&&	.	&&	.	&&	.	&&	.	&&	.	&&	.	&&	.	&&	.	&&	. && 10	\\
$\xten$	         & 	6	&&	2	&	$-$&2	&&	3	&	$-$&1	&&	.	&&	.	&	$-$&2	&&	.	&&	2	&&	.	&&	.	&&	.	&&	.	&&	.	&&	. && 15	\\
$\xeleven$	     & 	6	&&	2	&	$-$&2	&&	3	&	$-$&1	&&	.	&&	.	&&	2	&&	.	&	$-$&2	&&	.	&&	.	&&	.	&&	.	&&	.	&&	. && 15	\\
$\xtwelve$	     & 	9	&	$-$&3	&&	1	&&	.	&&	.	&&	.	&&	.	&	$-$&1	&&	1	&	$-$&1	&	$-$&3	&&	1	&&	.	&&	3	&	$-$&1	&&	. && 3	\\
$\xthirteen$	 & 	9	&	$-$&3	&&	1	&&	.	&&	.	&&	.	&&	.	&	$-$&1	&&	1	&	$-$&1	&&	3	&	$-$&1	&&	.	&	$-$&3	&&	1	&&	. && 6	\\
$\xfourteen$	 & 	9	&	$-$&3	&&	1	&&	.	&&	.	&&	.	&&	.	&&	1	&	$-$&1	&&	1	&	$-$&3	&&	1	&&	.	&	$-$&3	&&	1	&&	. && 3\\
$\xfifteen$	     & 	9	&	$-$&3	&&	1	&&	.	&&	.	&&	.	&&	.	&&	1	&	$-$&1	&&	1	&&	3	&	$-$&1	&&	.	&&	3	&	$-$&1	&&	. && 6	\\
$\xsixteen$	     & 	12	&&	4	&	$-$&4	&	$-$&3	&&	1	&&	.	&&	.	&&	.	&&	.	&&	.	&&	.	&&	.	&&	.	&&	.	&&	.	&&	. && 6	\\
    \hline\\[-0.4cm]\hline\\[-0.4cm]
  \end{tabular}
  }%scalebox
  \begin{flushleft}
    $^\text{a}$ %          
      Zero characters are labeled with ``.'' to enhance readability.\\
    $^\text{b}$ %      
      Class index. \\
    $^\text{c}$ %            
      Number of elements in class Cl.$[R]$.\\
    $^\text{d}$ %            
      Spin statistical weight.
  \end{flushleft}
\end{table*}

\clearpage

\section{$T_\text{d}$(M) character table}
\begin{table}[h!]
  \caption{%
    Characters and irrep decomposition 
    of the rotational functions of methane
    in the $T_\text{d}$(M) molecular symmetry group \cite{SaCsMa17}.
    \label{tab:td_chartab}
  }
\scalebox{0.88}{%
  \begin{tabular}{@{}l r@{}l r@{}l r@{}l r@{}l r@{}l c@{}}
  \cline{1-12} \\[-0.4cm]
  \cline{1-12} %\\[-0.4cm]
    & 
    \multicolumn{2}{c}{$E$} & 
    \multicolumn{2}{c}{(123)} & 
    \multicolumn{2}{c}{(14)(23)} & 
    \multicolumn{2}{c}{[(1423)]$^\ast$} & 
    \multicolumn{2}{c}{[(23)]$^\ast$} & Irreps \\
     & 
    \multicolumn{2}{c}{1} & 
    \multicolumn{2}{c}{8} & 
    \multicolumn{2}{c}{3} & 
    \multicolumn{2}{c}{6} & 
    \multicolumn{2}{c}{6} & \\        
  \cline{1-12} \\[-0.4cm]
    $j^{\text{M}}_k$
    & 
    \multicolumn{2}{c}{$2j^{\rm M}+1$} & 
    \multicolumn{2}{c}{%
\tiny
$\sum\limits_{m=-j^{\rm M}}^{j^{\rm M}} D^{j^{\rm M}\ast}_{m,m}(0,0,\frac{2\pi}{3})$
} & 
    \multicolumn{2}{c}{%
\tiny
$\sum\limits_{m=-j^{\rm M}}^{j^{\rm M}} D^{j^{\rm M}\ast}_{m,m}(\frac{\pi}{3},\pi,0)$
} & 
    \multicolumn{2}{c}{%
\tiny
      $\sum\limits_{m=-j^{\rm M}}^{j^{\rm M}} D^{j^{\rm M}\ast}_{m,m}(\frac{\pi}{6},\frac{\pi}{2},-\frac{\pi}{6})$
} & 
    \multicolumn{2}{c}{%
\tiny
$\sum\limits_{m=-j^{\rm M}}^{j^{\rm M}} D^{j^{\rm M}\ast}_{m,m}(0,\pi,0)$
} & \\
\cline{1-12} \\[-0.4cm]
    0 & 
   &1 &   
   &1 &
   &1 &
   &1 &
   &1 & A$_1$ \\
    1 & 
   &3 & 
   &0 &
$-$&1 &
   &1 &
$-$&1 & F$_1$  \\
    2 & 
   &5 & 
$-$&1 &
   &1 &
$-$&1 &
   &1 & $\text{E}\oplus \text{F}_2$ \\
    3 & 
   &7 & 
   &1 &
$-$&1 &
$-$&1 &
$-$&1 & $\text{A}_2\oplus\text{F}_1\oplus\text{F}_2$ \\
    4 & 
   &9 & 
   &0 &
   &1 &
   &1 &
   &1 & $\text{A}_1\oplus\text{E}\oplus\text{F}_1\oplus\text{F}_2$ \\
  5 & 
  &11 & 
$-$&1 &
$-$&1 &
   &1 &
$-$&1 & $\text{E}\oplus 2\text{F}_1\oplus \text{F}_2$ \\
  \cline{1-12} \\[-0.4cm]
  \cline{1-12} 
  \end{tabular}  
}
\end{table}

\clearpage
\begin{figure}[h!]
    \centering
    \includegraphics[width=0.9\textwidth]{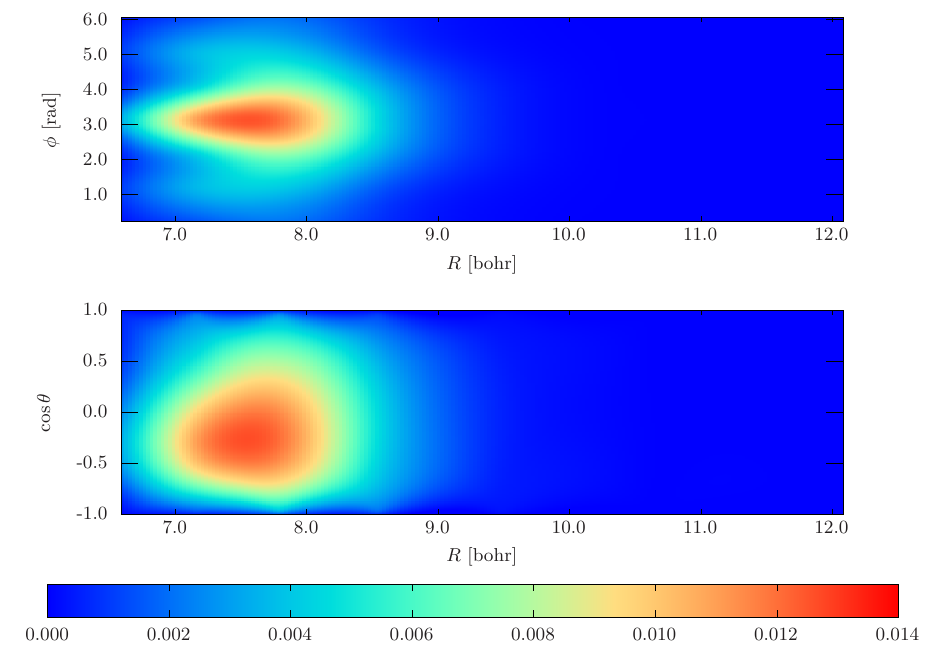}
    \caption{%
    2D cut of the wave function of the ground state, $[1,0]$, of the ortho-meta spin isomer corresponding to the
    J0.2 (from the 6-fold degenerate J0.2--7) state  (Table~\ref{tab:mm_states1}). }
    \label{fig:om_wf}
\end{figure}
\begin{figure}[h!]
    \centering
     \includegraphics[width=0.9\textwidth]{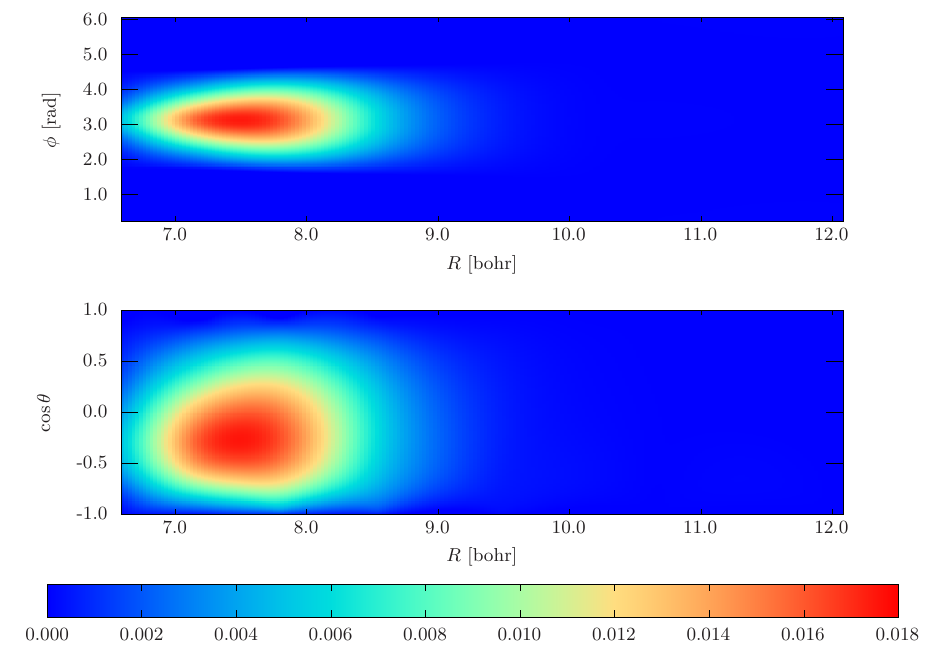}
    \caption{%
    2D cut of the wave function of the ground state, $[1,1]$, of the ortho-ortho spin isomer corresponding to the
    J0.8 (from the 9-fold degenerate J0.8--16) state  (Table~\ref{tab:mm_states1}). }
    \label{fig:oo_wf}
\end{figure}
\begin{figure}[h!]
    \centering
    \includegraphics[width=0.9\textwidth]{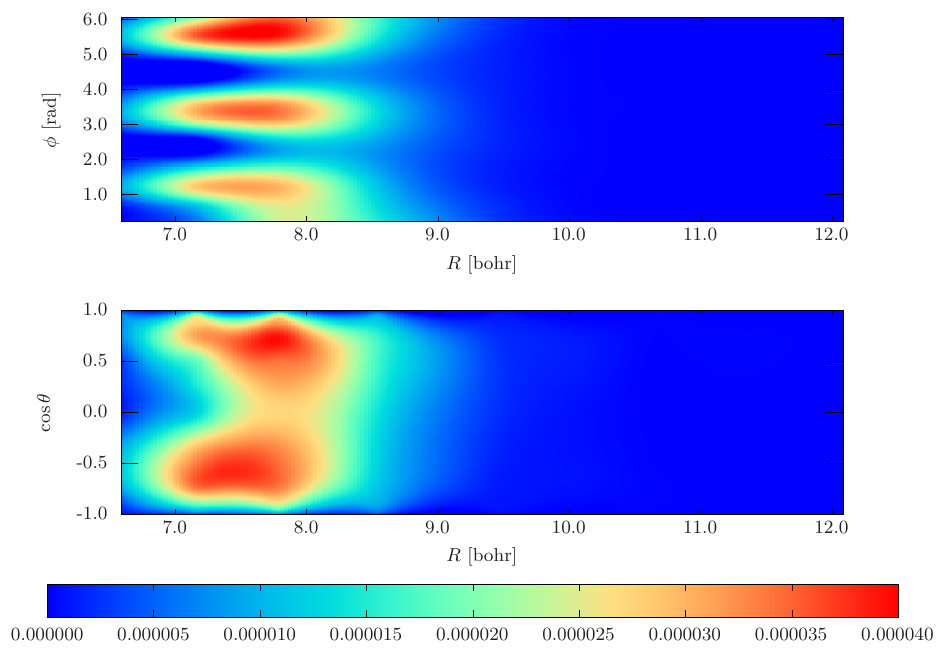}
    \caption{%
    2D cut of the wave function of the ground state, $[2,0]$, of the meta-para spin isomer corresponding to the
    J1.83 (from the 4-fold degenerate J1.83--86) state  (Table~\ref{tab:mm_states1}). }
    \label{fig:mp_wf}
\end{figure}
\begin{figure}[h!]
    \centering
    \includegraphics[width=0.9\textwidth]{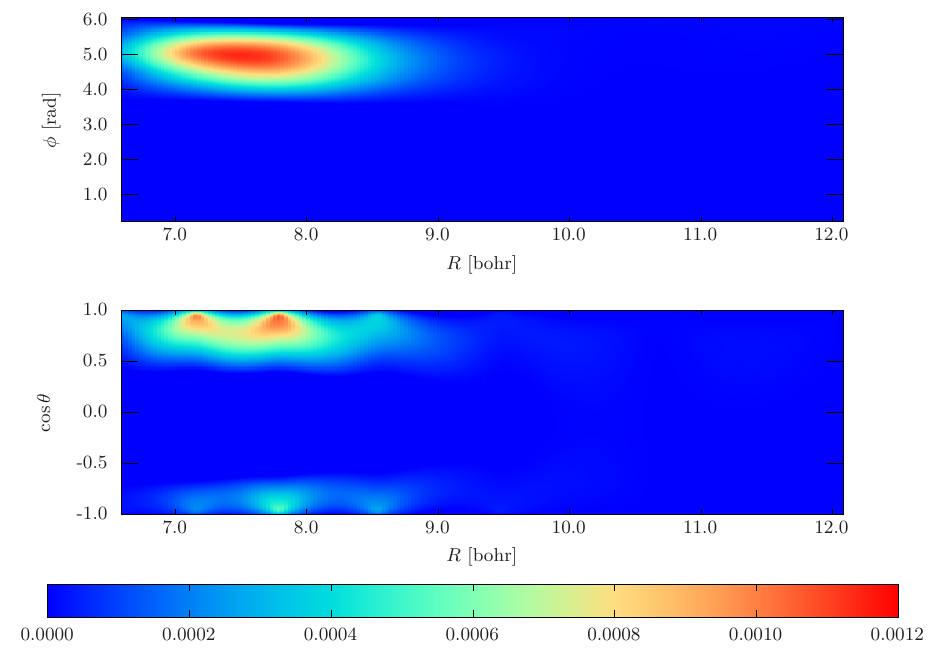}
    \caption{%
    2D cut of the wave function of the ground state, $[2,1]$, of the ortho-para spin isomer corresponding to the
    J1.93 (from the 12-fold degenerate J1.93--104) state  (Table~\ref{tab:mm_states1}). }
    \label{fig:op_wf}
\end{figure}

\begin{table}[H]
\caption{%
$U^{(\Omega)}_{\omega\sigma,\alpha}$ matrix elements for $\Omega=2$. 
  \label{tab:u2}  
}
\begin{center}
\begin{tabular}{ c@{\ \ \ } c@{\ \ \ } c@{\ \ \ } c@{\ \ \ } c@{\ \ \ } c@{\ \ \ } c@{\ \ \ } c@{\ \ \ } c@{\ \ \ } c@{\ \ \ } c@{\ \ \ } c@{\ \ \ } c@{\ \ \ }}
\hline\\[-0.35cm]
\hline\\[-0.35cm]
($\omega, \sigma$) &$\alpha$:& $xx$ && $xy$ && $xz$ && $yy$ && $yz$ && $zz$\\
\hline\\[-0.20cm]
(0,0)  && $-\frac{1}{\sqrt{3}}$ && 0 && 0 && $-\frac{1}{\sqrt{3}}$ && 0 && $-\frac{1}{\sqrt{3}}$ \\[0.25cm]
(2,$-$2)  && $\frac{1}{2}$ && $-\frac{\rm i}{2}$ && 0 && $-\frac{1}{2}$ && 0 && 0 \\[0.25cm]
(2,$-$1)  && 0 && 0 && $\frac{1}{2}$ && 0 && $-\frac{\rm i}{2}$ && 0 \\[0.25cm]
(2,0)   && $-\frac{1}{\sqrt{6}}$ && 0 && 0 && $-\frac{1}{\sqrt{6}}$ && 0 && $\sqrt{\frac{2}{3}}$ \\[0.25cm]
(2,1)  && 0 && 0 && $-\frac{1}{2}$ && 0 && $-\frac{\rm i}{2}$ && 0 \\[0.25cm]
(2,2)  && $\frac{1}{2}$ && $\frac{\rm i}{2}$ && 0 && $-\frac{1}{2}$ && 0 && 0 \\[0.15cm]
\hline\\[-0.35cm]
\hline\\[-0.35cm]
\end{tabular}
\end{center}
\begin{flushleft}
\end{flushleft}
\end{table}
%%%%%%%%%%
%\vspace{-1.5cm}
%%%%%%%%%%
\begin{table}[H]
\caption{%
$[U^{(\Omega)}_{\omega\sigma,\alpha}]^{-1}$ matrix elements for $\Omega=2$. 
  \label{tab:u2_inv}  
}
\begin{center}
\begin{tabular}{ c@{\ \ \ } c@{\ \ \ } c@{\ \ \ } c@{\ \ \ } c@{\ \ \ } c@{\ \ \ } c@{\ \ \ } c@{\ \ \ } c@{\ \ \ } c@{\ \ \ } c@{\ \ \ } c@{\ \ \ } c@{\ \ \ }}
\hline\\[-0.35cm]
\hline\\[-0.3cm]
$\alpha$ &($\omega, \sigma$):& (0,0) && (2,$-$2) && (2,$-$1) && (2,0) && (2,1) && (2,2) \\
\hline\\[-0.20cm]
$xx$  && $-\frac{1}{\sqrt{3}}$ && $\frac{1}{2}$ && 0 && $-\frac{\text{1}}{\sqrt{6}}$ && 0 && $\frac{1}{2}$ \\[0.25cm]
$xy$  && 0 && $\frac{\rm i}{2}$ && 0 && 0 && 0 && $-\frac{\rm i}{2}$ \\[0.25cm]
$xz$  && 0 && 0 && $\frac{1}{2}$ && 0 && $\frac{1}{2}$ && 0 \\[0.25cm]
$yy$  && $-\frac{1}{\sqrt{3}}$ && $-\frac{1}{2}$ && 0 && $-\frac{\text{1}}{\sqrt{6}}$ && 0 && $-\frac{1}{2}$ \\[0.25cm]
$yz$  && 0 && 0 && $\frac{\rm i}{2}$ && 0 && $\frac{\rm i}{2}$ && 0 \\[0.25cm]
$zz$  && $-\frac{1}{\sqrt{3}}$ && 0 && 0 && $\sqrt{\frac{2}{3}}$ && 0 && 0 \\[0.15cm]
\hline\\[-0.35cm]
\hline\\[-0.35cm]
\end{tabular}
\end{center}
\begin{flushleft}
\end{flushleft}
\end{table}

\begin{table}[H]
\caption{%
  Computed $J=0$ and $J=1$ rovibrational states of the \metmet. The energies are referenced to the ZPVE (94.25 \cm).
  %\\[-0.5cm]
  %
  \label{tab:mm_states1}  
}
\singlespacing
\centering
\scalebox{0.80}{
\begin{tabular}{@{}r@{\ \ } r@{\ \ } l@{\ \ } r@{\ \ } c@{\ \ } r@{\ } r@{\ \ } l@{\ \ } r@{\ \ } c@{}}
\hline\\[-0.35cm]
\hline\\[-0.35cm]
Label & 
\multicolumn{1}{c}{$\tilde\nu$ [\cm]}	&
\multicolumn{1}{l}{Assignment}&
\multicolumn{1}{c}{Symm.}
&& Label & 
\multicolumn{1}{c}{$\tilde\nu$ [\cm]}	&
\multicolumn{1}{c}{Assignment}&
\multicolumn{1}{c}{Symm.}
\\
\hline\\[-0.35cm]
 $J0.n$ &&&&& $J1.n$ \\
 \hline\\[-0.35cm]
  1           &   0.00    &   $   [[0,0]_0,0]_{00}            $   &   $\xone$      &  &1          &   0.24    &   $   [[0,0]_0,1]_{10}            $   &   $\xtwo$         \\
 2--7        &   8.99    &   $   [[1,0]_1,1]_{00}            $   &   $\xeleven$   &  &2--7       &   9.12    &   $   [[1,0]_1,0]_{10}            $   &   $\xten$         \\
 8--16       &   18.04   &   $   [[1,1]_L,L]_{00}\ (L=0,2)    $   &   $\xfifteen$  &  &8--13      &   9.80    &   $   [[1,0]_1,1]_{10}            $   &   $\xeleven$      \\
 17--25      &   19.32   &   $   [[1,1]_1,1]_{00}            $   &   $\xtwelve$   &  &14--19     &   9.91    &   $   [[1,0]_1,2]_{10}            $   &   $\xten$         \\
 26--34      &   19.37   &   $   [[1,1]_L,L]_{00}\ (L=0,2)    $   &   $\xfifteen$  &  &20--28     &   18.10   &   $   [[1,1]_0,1]_{10}            $   &   $\xtwelve$      \\
 35--40      &   31.43   &   $   [[2,0]_2,2]_{00}            $   &   $\xeleven$   &  &29--37     &   18.65   &   $   [[1,1]_1,0]_{10}            $   &   $\xfifteen$     \\
 41--44      &   31.73   &   $   [[2,0]_2,2]_{00}            $   &   $\xnine$     &  &38--46     &   18.74   &   $   [[1,1]_2,1]_{10}            $   &   $\xtwelve$      \\
 45--56      &   32.17   &   $   [[2,1]_L,L]_{00}\ (L=1,2,3)  $   &   $\xsixteen$  &  &47--55     &   18.90   &   $   [[1,1]_1,2]_{10}            $   &   $\xfifteen$     \\
 57          &   32.93   &   $   [[0,0]_0,0]_{00}            $   &   $\xone$      &  &56--64     &   18.95   &   $   [[1,1]_1,1]_{10}            $   &   $\xtwelve$      \\
 58--66      &   38.44   &   $   [[2,1]_3,3]_{00}            $   &   $\xfourteen$ &  &65--73     &   19.64   &   $   [[1,1]_2,2]_{10}            $   &   $\xfifteen$     \\
 67--75      &   38.88   &   $   [[2,1]_1,1]_{00}            $   &   $\xfifteen$  &  &74--82     &   19.74   &   $   [[1,1]_2,3]_{10}            $   &   $\xtwelve$      \\
 76--84      &   40.47   &   $   [[2,1]_2,2]_{00}            $   &   $\xtwelve$   &  &83--86     &   22.52   &   $   [[2,0]_2,L]_{10}\ (L=1,2,3)  $   &   $\xnine$        \\
 85--96      &   40.52   &   $   [[2,1]_L,L]_{00}\ (L=1,2,3)  $   &   $\xsixteen$  &  &87--92     &   31.29   &   $   [[2,0]_2,1]_{10}            $   &   $\xten$         \\
 97--105     &   42.19   &   $   [[2,1]_2,2]_{00}            $   &   $\xtwelve$   &  &93--104    &   31.67   &   $   [[2,1]_j,\Lambda]_{10}\ (\mathrm{mixed})    $   &   $\xsixteen$     \\
 106--114    &   42.43   &   $   [[2,1]_1,1]_{00}            $   &   $\xfourteen$ &  &105--110   &   31.82   &   $   [[2,0]_2,2]_{10}            $   &   $\xeleven$      \\  
 115--123    &   43.57   &   $   [[2,1]_3,3]_{00}            $   &   $\xfifteen$  &  &111--114   &   31.97   &   $   [[2,0]_2,L]_{10}\ (L=1,2,3)  $   &   $\xnine$        \\     
 124--129    &   45.14   &   $   [[1,0]_1,1]_{00}            $   &   $\xeleven$   &  &115--120   &   32.20    &   $   [[2,0]_2,3]_{10}            $   &   $\xten$         \\      
             &           &                                       &                &  &121--132   &   32.45   &   $   [[2,1]_j,\Lambda]_{10}\  (\mathrm{mixed})    $   &   $\xsixteen$     \\
             &           &                                       &                &  &133        &   33.15   &   $   [[0,0]_0,1]_{10}            $   &   $\xtwo$         \\
\hline\\[-0.35cm]
\hline
\end{tabular}
}
\end{table}

\begin{table}[H]
\caption{%
  Computed $J=2$ and $J=3$ rovibrational states of the \metmet. The energies are referenced to the ZPVE (94.25 \cm).
  \label{tab:mm_states2}  
}
\singlespacing
\centering
\scalebox{0.78}{
\begin{tabular}{@{}r@{\ \ } r@{\ \ } l@{\ \ } r@{\ \ } c@{\ \ } r@{\ } r@{\ \ } l@{\ \ } r@{\ \ } c@{}}
\hline\\[-0.35cm]
\hline\\[-0.35cm]
Label & 
\multicolumn{1}{c}{$\tilde\nu$ [\cm]}	&
\multicolumn{1}{l}{Assignment}&
\multicolumn{1}{c}{Symm.}
&& Label & 
\multicolumn{1}{c}{$\tilde\nu$ [\cm]}	&
\multicolumn{1}{c}{Assignment}&
\multicolumn{1}{c}{Symm.}
\\
\hline\\[-0.35cm]
 $J2.n$ &&&&& $J3.n$ \\
 \hline\\[-0.35cm]
 1   &   0.73    &   $   [[0,0]_0,2]_{20}    $   &   $\xone$ &   &   1   &   1.49    &   $   [[0,0]_0,3]_{30}    $   &   $\xtwo$ \\
 2--7    &   9.44    &   $   [[1,0]_1,1]_{20}    $   &   $\xeleven$  &   &   2--7    &   10.00   &   $   [[1,0]_1,2]_{30}    $   &   $\xten$ \\
 8--13   &   10.29   &   $   [[1,0]_1,2]_{20}    $   &   $\xten$ &   &   8--13   &   11.02   &   $   [[1,0]_1,3]_{30}    $   &   $\xeleven$  \\
 14--19  &   10.55   &   $   [[1,0]_1,3]_{20}    $   &   $\xeleven$  &   &   14--19  &   11.46   &   $   [[1,0]_1,4]_{30}    $   &   $\xten$ \\
 20--28  &   18.28   &   $   [[1,1]_2,0]_{20}    $   &   $\xfifteen$ &   &   20--28  &   18.66   &   $   [[1,1]_2,1]_{30}    $   &   $\xtwelve$  \\
 29--37  &   19.01   &   $   [[1,1]_1,1]_{20}    $   &   $\xtwelve$  &   &   29--37  &   19.58   &   $   [[1,1]_2,2]_{30}    $   &   $\xfifteen$ \\
 38--46  &   19.10   &   $   [[1,1]_2,1]_{20}    $   &   $\xtwelve$  &   &   38--46  &   19.61   &   $   [[1,1]_1,2]_{30}    $   &   $\xfifteen$ \\
 47--55  &   19.22   &   $   [[1,1]_0,2]_{20}    $   &   $\xfifteen$ &   &   47--55  &   19.95   &   $   [[1,1]_0,3]_{30}    $   &   $\xtwelve$  \\
 56--64  &   19.31   &   $   [[1,1]_1,2]_{20}    $   &   $\xfifteen$ &   &   56--64  &   20.00   &   $   [[1,1]_1,3]_{30}    $   &   $\xtwelve$  \\
 65--73  &   19.80   &   $   [[1,1]_2,2]_{20}    $   &   $\xfifteen$ &   &   65--73  &   20.57   &   $   [[1,1]_2,3]_{30}    $   &   $\xtwelve$  \\
 74--82  &   19.92   &   $   [[1,1]_1,3]_{20}    $   &   $\xtwelve$  &   &   74--82  &   20.58   &   $   [[1,1]_2,4]_{30}    $   &   $\xfifteen$ \\
 83--91  &   20.27   &   $   [[1,1]_2,3]_{20}    $   &   $\xtwelve$  &   &   83--91  &   21.16   &   $   [[1,1]_1,4]_{30}    $   &   $\xfifteen$ \\
 92--100 &   20.50   &   $   [[1,1]_2,4]_{20}    $   &   $\xfifteen$ &   &   92--100 &   21.58   &   $   [[1,1]_2,5]_{30}    $   &   $\xtwelve$  \\
 101--104    &   22.98   &   $   [[2,0]_2,L]_{20}\ (L=0,1,2,3,4)  $   &   $\xnine$    &   &   101--104    &   23.66   &   $   [[2,0]_2,L]_{30}\ (L=1,2,4)  $   &   $\xnine$    \\
 105--108    &   26.68  &   $   [[2,0]_2,L]_{20}\ (L=0,1,2,3,4)  $   &   $\xnine$    &   &   105--108    &   27.40   &   $   [[2,0]_2,L]_{30}\ (L=2,3)    $   &   $\xnine$    \\
 109--114    &   29.86   &   $   [[2,0]_2,0]_{20}    $   &   $\xeleven$  &   &   109--114    &   30.46   &   $   [[2,0]_2,1]_{30}    $   &   $\xten$ \\
 115--120    &   29.87   &   $   [[2,0]_2,1]_{20}    $   &   $\xten$ &   &   115--120    &   30.50   &   $   [[2,0]_2,2]_{30}    $   &   $\xeleven$  \\
 121--126    &   31.54   &   $   [[2,0]_2,2]_{20}    $   &   $\xeleven$  &   &   121--126    &   32.03   &   $   [[2,0]_2,3]_{30}    $   &   $\xten$ \\
 127--138    &   31.93   &   $   [[2,1]_j,\Lambda]_{20}\ (\mathrm{mixed})    $   &   $\xsixteen$ &   &   127--138    &   32.40   &   $   [[2,1]_j,\Lambda]_{30}\ (\mathrm{mixed})    $   &   $\xsixteen$ \\
 139--144    &   32.38   &   $   [[2,0]_2,3]_{20}    $   &   $\xten$ &   &   139--142    &   33.14   &   $   [[2,0]_2,L]_{30}\ (L=1,3,5)  $   &   $\xnine$    \\
 145--148    &   32.42   &   $   [[2,0]_2,4]_{20}    $   &   $\xnine$    &   &   143--148    &   33.20   &   $   [[2,0]_2,4]_{30}    $   &   $\xeleven$  \\
 149--160    &   32.54   &   $   [[2,1]_j,\Lambda]_{20}\ (\mathrm{mixed})$   &   $\xsixteen$ &   &   149--160    &   33.27   &   $   [[2,1]_j,\Lambda]_{30}\ (\mathrm{mixed})$   &   $\xsixteen$ \\
 161--172    &   33.04   &   $   [[2,1]_j,\Lambda]_{20}\ (\mathrm{mixed})$   &   $\xsixteen$ &   &   161 &   33.85   &   $   [[0,0]_0,3]_{30}    $   &   $\xtwo$ \\
 173--178    &   33.06   &   $   [[2,0]_2,4]_{20}    $   &   $\xeleven$  &   &   162--173    &   33.87   &   $   [[2,1]_j,\Lambda]_{30}\ (\mathrm{mixed})    $   &   $\xsixteen$ \\
 179 &   33.19   &   $   [[0,0]_0,2]_{20}    $   &   $\xone$ &   &   174--179    &   34.13   &   $   [[2,0]_2,5]_{30}    $   &   $\xten$ \\
  180--191    &   35.72   &   $   [[2,1]_j,\Lambda]_{20}\ (\mathrm{mixed})$   &   $\xsixteen$ &   &   180--191    &   36.12   &   $   [[2,1]_j,\Lambda]_{30}\ (\mathrm{mixed})$   &   $\xsixteen$ \\
 192--195    &   36.02    &   $   [[2,0]_2,L]_{20}\ (L=0,1,2)  $   &   $\xnine$    &   &   192--195    &   36.75   &   $   [[2,0]_2,L]_{30}\ (L=1,2,3)  $   &   $\xnine$    \\
 196--207    &   36.74   &   $   [[2,1]_j,\Lambda]_{20}\ (\mathrm{mixed})$   &   $\xsixteen$ &   &   196--207    &   36.81   &   $   [[2,1]_j,\Lambda]_{30}\ (\mathrm{mixed})$   &   $\xsixteen$ \\
208--216    &   38.44   &   $   [[2,1]_j,\Lambda]_{20}\ (\mathrm{mixed})    $   &   $\xfifteen$ &   &   208--219    &   37.59   &   $   [[2,1]_j,\Lambda]_{30}\ (\mathrm{mixed})    $   &   $\xsixteen$ \\
 217--225    &   38.46   &   $   [[2,1]_j,\Lambda]_{20}\ (\mathrm{mixed})    $   &   $\xtwelve$  &   &   220--228    &   38.84   &   $   [[2,1]_j,\Lambda]_{30}\ (\mathrm{mixed})    $   &   $\xtwelve$  \\
 226--234    &   38.53   &   $   [[2,1]_j,\Lambda]_{20}\ (\mathrm{mixed})    $   &   $\xfifteen$ &   &   229--237    &   38.84   &   $   [[2,1]_j,\Lambda]_{30}\ (\mathrm{mixed})    $   &   $\xtwelve$  \\
 235--243    &   38.60   &   $   [[2,1]_j,\Lambda]_{20}\ (\mathrm{mixed})    $   &   $\xtwelve$  &   &   238--246    &   38.92   &   $   [[2,1]_j,\Lambda]_{30}\ (\mathrm{mixed})    $   &   $\xfifteen$ \\
 244--252    &   38.84   &   $   [[2,1]_j,\Lambda]_{20}\ (\mathrm{mixed})    $   &   $\xfifteen$ &   &   247--255    &   38.97   &   $   [[2,1]_j,\Lambda]_{30}\ (\mathrm{mixed})    $   &   $\xfifteen$ \\
 253--261    &   39.22   &   $   [[2,1]_j,\Lambda]_{20}\ (\mathrm{mixed})    $   &   $\xfifteen$ &   &   259--264    &   39.38   &   $   [[2,1]_j,\Lambda]_{30}\ (\mathrm{mixed})    $   &   $\xtwelve$  \\
 262--270    &   39.32   &   $   [[2,1]_j,\Lambda]_{20}\ (\mathrm{mixed})    $   &   $\xfifteen$ &   &   265--273    &   39.55   &   $   [[2,1]_j,\Lambda]_{30}\ (\mathrm{mixed})    $   &   $\xtwelve$  \\
 271--279    &   39.35   &   $   [[2,1]_j,\Lambda]_{20}\ (\mathrm{mixed})    $   &   $\xtwelve$  &   &   274--282    &   39.56   &   $   [[2,1]_j,\Lambda]_{30}\ (\mathrm{mixed})    $   &   $\xfifteen$ \\
 280--288    &   39.80   &   $   [[2,1]_j,\Lambda]_{20}\ (\mathrm{mixed})    $   &   $\xtwelve$  &   &   283--291    &   39.75   &   $   [[2,1]_j,\Lambda]_{30}\ (\mathrm{mixed})    $   &   $\xtwelve$  \\
 289--297    &   39.99   &   $   [[2,1]_j,\Lambda]_{20}\ (\mathrm{mixed})    $   &   $\xtwelve$  &   &   292--300    &   40.01   &   $   [[2,1]_j,\Lambda]_{30}\ (\mathrm{mixed})    $   &   $\xtwelve$  \\
 298--306    &   40.13   &   $   [[2,1]_j,\Lambda]_{20}\ (\mathrm{mixed})    $   &   $\xfifteen$ &   &   301--309    &   40.04   &   $   [[2,1]_j,\Lambda]_{30}\ (\mathrm{mixed})    $   &   $\xfifteen$ \\
 307--315    &   40.21   &   $   [[2,1]_j,\Lambda]_{20}\ (\mathrm{mixed})    $   &   $\xfifteen$ &   &   310--318    &   40.06   &   $   [[2,1]_j,\Lambda]_{30}\ (\mathrm{mixed})    $   &   $\xtwelve$  \\
 316--324    &   40.43   &   $   [[2,1]_j,\Lambda]_{20}\ (\mathrm{mixed})    $   &   $\xtwelve$  &   &   319--327    &   40.14   &   $   [[2,1]_j,\Lambda]_{30}\ (\mathrm{mixed})    $   &   $\xfifteen$ \\
 325--333    &   40.64   &   $   [[2,1]_j,\Lambda]_{20}\ (\mathrm{mixed})    $   &   $\xfifteen$ &   &   328--336    &   40.57   &   $   [[2,1]_j,\Lambda]_{30}\ (\mathrm{mixed})    $   &   $\xfifteen$ \\
 334--345    &   40.96   &   $   [[2,1]_j,\Lambda]_{20}\ (\mathrm{mixed})    $   &   $\xsixteen$ &   &   337--345    &   40.58   &   $   [[2,1]_j,\Lambda]_{30}\ (\mathrm{mixed})    $   &   $\xfifteen$ \\
 346--349    &   41.15   &   $   [[2,0]_2,L]_{20}\ (L=3,4)    $   &   $\xnine$    &   &   346--354    &   40.80   &   $   [[2,1]_j,\Lambda]_{30}\ (\mathrm{mixed})$   &   $\xtwelve$  \\
 350--358    &   41.20  &   $   [[2,1]_j,\Lambda]_{20}\ (\mathrm{mixed})$   &   $\xtwelve$  &   &   355--363    &   40.97   &   $   [[2,1]_j,\Lambda]_{30}\ (\mathrm{mixed})$   &   $\xfifteen$ \\
     &       &   $       $   &       &   &   364--372    &   41.12   &   $   [[2,1]_j,\Lambda]_{30}\ (\mathrm{mixed})$   &   $\xtwelve$  \\
     &       &   $       $   &       &   &   373--381    &   41.38   &   $   [[2,1]_j,\Lambda]_{30}\ (\mathrm{mixed})$   &   $\xtwelve$  \\
\hline\\[-0.35cm]
\hline
\end{tabular}
}
\end{table}

\begin{table}[H]
\caption{%
  Computed $J=4,5,$ and $6$ rovibrational states of the \metmet. The energies are referenced to the ZPVE (94.25 \cm).
  \label{tab:mm_states3}  
}
\singlespacing
\centering
\scalebox{0.78}{
\begin{tabular}{@{}r@{\ \ } r@{\ \ } l@{\ \ } r@{\ \ } c@{\ \ } r@{\ } r@{\ \ } l@{\ \ } r@{\ \ } c@{}}
\hline\\[-0.35cm]
\hline\\[-0.35cm]
Label & 
\multicolumn{1}{c}{$\tilde\nu$[\cm]}	&
\multicolumn{1}{l}{Assignment}&
\multicolumn{1}{c}{Symm.}
&& Label & 
\multicolumn{1}{c}{$\tilde\nu$[\cm]}	&
\multicolumn{1}{c}{Assignment}&
\multicolumn{1}{c}{Symm.}
\\
\hline\\[-0.35cm]
 $J4.n$ &&&&& $J5.n$ \\
 \hline\\[-0.35cm]
 1   &   2.43    &   $   [[0,0]_0,4]_{40}    $   &   $\xone$ &   &   1   &   3.65    &   $   [[0,0]_0,5]_{50}    $   &   $\xtwo$ \\
 2--7    &   10.79   &   $   [[1,0]_1,3]_{40}    $   &   $\xeleven$  &   &   2--7    &   11.82   &   $   [[1,0]_1,4]_{50}    $   &   $\xten$ \\
 8--13   &   12.00   &   $   [[1,0]_1,4]_{40}    $   &   $\xten$ &   &   8--13   &   13.21   &   $   [[1,0]_1,5]_{50}    $   &   $\xeleven$  \\
 14--19  &   12.61   &   $   [[1,0]_1,5]_{40}    $   &   $\xeleven$  &   &   14--19  &   14.01   &   $   [[1,0]_1,6]_{50}    $   &   $\xten$ \\
 20--28  &   19.25   &   $   [[1,1]_2,2]_{40}    $   &   $\xfifteen$ &   &   20--28  &   20.09   &   $   [[1,1]_2,3]_{50}    $   &   $\xtwelve$  \\
 29--37  &   20.38   &   $   [[1,1]_1,3]_{40}    $   &   $\xtwelve$  &   &   29--37  &   21.39   &   $   [[1,1]_1,4]_{50}    $   &   $\xfifteen$ \\
 38--46  &   20.38   &   $   [[1,1]_2,3]_{40}    $   &   $\xtwelve$  &   &   38--46  &   21.42   &   $   [[1,1]_2,4]_{50}    $   &   $\xfifteen$ \\
 47--55  &   20.92   &   $   [[1,1]_0,4]_{40}    $   &   $\xfifteen$ &   &   47--55  &   22.13   &   $   [[1,1]_0,5]_{50}    $   &   $\xtwelve$  \\
 56--64  &   20.95   &   $   [[1,1]_1,4]_{40}    $   &   $\xfifteen$ &   &   56--64  &   22.15   &   $   [[1,1]_1,5]_{50}    $   &   $\xtwelve$  \\
 65--73  &   21.57   &   $   [[1,1]_2,4]_{40}    $   &   $\xfifteen$ &   &   65--73  &   22.80   &   $   [[1,1]_2,5]_{50}    $   &   $\xtwelve$  \\
 74--82  &   22.05   &   $   [[1,1]_2,5]_{40}    $   &   $\xtwelve$  &   &   74--82  &   23.47   &   $   [[1,1]_2,6]_{50}    $   &   $\xfifteen$ \\
 83--91  &   22.30   &   $   [[1,1]_1,5]_{40}    $   &   $\xtwelve$  &   &   83--91  &   23.70   &   $   [[1,1]_1,6]_{50}    $   &   $\xfifteen$ \\
 92--100 &   22.92   &   $   [[1,1]_2,6]_{40}    $   &   $\xfifteen$ &   &   92--100 &   24.51   &   $   [[1,1]_2,7]_{50}    $   &   $\xtwelve$  \\
  101--104    &   24.56   &   $   [[2,0]_2,L]_{40}\ (L=2,3,4,5,6)  $   &   $\xnine$    &   &   101--104    &   25.69   &   $   [[2,0]_2,L]_{50}\ (L=3,6)    $   &   $\xnine$    \\
 105--108    &   28.36   &   $   [[2,0]_2,L]_{40}\ (L=2,3,4,5,6)  $   &   $\xnine$    &   &   105--108    &   29.56   &   $   [[2,0]_2,L]_{40}\ (L=4,5)    $   &   $\xnine$    \\
 109--114    &   31.25   &   $   [[2,0]_2,2]_{40}    $   &   $\xeleven$  &   &   109--114    &   32.22   &   $   [[2,0]_2,3]_{50}    $   &   $\xten$ \\
 115--120    &   31.34   &   $   [[2,0]_2,3]_{40}    $   &   $\xten$ &   &   115--120    &   32.41   &   $   [[2,0]_2,4]_{50}    $   &   $\xeleven$  \\
 121--126    &   32.80   &   $   [[2,0]_2,4]_{40}    $   &   $\xeleven$  &   &   121--126    &   33.87   &   $   [[2,0]_2,5]_{50}    $   &   $\xten$ \\
 127--138    &   33.10   &   $   [[2,1]_j,\Lambda]_{40}\ (\mathrm{mixed})    $   &   $\xsixteen$ &   &   127--138    &   34.03   &   $   [[2,1]_j,\Lambda]_{50}\ (\mathrm{mixed})    $   &   $\xsixteen$ \\
 139--142    &   34.10   &   $   [[2,0]_2,L]_{40}\ (L=2,4,6)  $   &   $\xnine$    &   &   139--142    &   35.31   &   $   [[2,0]_2,7]_{50}    $   &   $\xnine$    \\
 143--154    &   34.20   &   $   [[2,1]_j,\Lambda]_{40}\ (\mathrm{mixed})    $   &   $\xsixteen$ &   &       &       &   $       $   &       \\[-0.1cm]
 155--160    &   34.29   &   $   [[2,0]_2,5]_{40}    $   &   $\xten$ &   &    \raisebox{0.2cm}{$J6.n$}    &      &   $       $   &       \\[-0.15cm]
 \cline{5-9}\\[-0.35cm]
 161 &   34.74   &   $   [[0,0]_0,4]_{40}    $   &   $\xone$ &   &   1   &   5.10    &   $   [[0,0]_0,6]_{60}    $   &   $\xone$ \\
 162--173    &   34.94   &   $   [[2,1]_j,\Lambda]_{40}\ (\mathrm{mixed})    $   &   $\xsixteen$ &   &   2--7    &   13.08   &   $   [[1,0]_1,5]_{60}    $   &   $\xeleven$  \\
 174--179    &   35.42   &   $   [[2,0]_2,5]_{40}    $   &   $\xeleven$  &   &   8--13   &   14.68   &   $   [[1,0]_1,6]_{60}    $   &   $\xten$ \\
 180--191    &   36.86   &   $   [[2,1]_j,\Lambda]_{40}\ (\mathrm{mixed})    $   &   $\xsixteen$ &   &   14--19  &   15.65   &   $   [[1,0]_1,7]_{60}    $   &   $\xeleven$  \\
 192--195    &   37.73   &   $   [[2,0]_2,L]_{40}\ (L=3,4,5)  $   &   $\xnine$    &   &   20--28  &   21.16   &   $   [[1,1]_2,4]_{60}    $   &   $\xfifteen$ \\
 196-207 &   37.84   &   $   [[2,1]_j,\Lambda]_{40}$ (mixed)   &   $\xsixteen$ &   &   29--37  &   22.65   &   $   [[1,1]_1,5]_{60}    $   &   $\xtwelve$  \\
 208--219    &   38.71   &   $   [[2,1]_j,\Lambda]_{40}$ (mixed)   &   $\xsixteen$ &   &   38--46  &   22.69   &   $   [[1,1]_2,5]_{60}    $   &   $\xtwelve$  \\
 220--228    &   39.38   &   $   [[2,1]_j,\Lambda]_{40}$ (mixed)   &   $\xfifteen$ &   &   47--55  &   23.58   &   $   [[1,1]_1,6]_{60}    $   &   $\xfifteen$ \\
 229--237    &   39.43   &   $   [[2,1]_j,\Lambda]_{40}$ (mixed)   &   $\xfifteen$ &   &   56--64  &   23.60   &   $   [[1,1]_0,6]_{60}    $   &   $\xfifteen$ \\
 238--246    &   39.59   &   $   [[2,1]_j,\Lambda]_{40}$ (mixed)   &   $\xtwelve$  &   &   65--73  &   24.27   &   $   [[1,1]_2,6]_{60}    $   &   $\xfifteen$ \\
 247--255    &   39.61   &   $   [[2,1]_j,\Lambda]_{40}$ (mixed)   &   $\xtwelve$  &   &   74--82  &   25.13   &   $   [[1,1]_2,7]_{60}    $   &   $\xtwelve$  \\
 256--264    &   40.14   &   $   [[2,1]_j,\Lambda]_{40}$ (mixed)   &   $\xfifteen$ &   &   83--91  &   25.34   &   $   [[1,1]_1,7]_{60}    $   &   $\xtwelve$  \\
 265--273    &   40.42   &   $   [[2,1]_j,\Lambda]_{40}$ (mixed)   &   $\xfifteen$ &   &   92--100 &   26.35   &   $   [[1,1]_2,8]_{60}    $   &   $\xfifteen$ \\
 274--282    &   40.43   &   $   [[2,1]_j,\Lambda]_{40}$ (mixed)   &   $\xtwelve$  &   &   101--104    &   27.04   &   $   [[2,0]_2,2]_{60}    $   &   $\xnine$    \\
 283--291    &   40.52   &   $   [[2,1]_j,\Lambda]_{40}$ (mixed)   &   $\xfifteen$ &   &   105--108    &   31.00   &   $   [[2,0]_2,L]_{60}\ (L=3,4)    $   &   $\xnine$    \\
 292--300    &   40.84   &   $   [[2,1]_j,\Lambda]_{40}$ (mixed)   &   $\xfifteen$ &   &   109--114    &   33.39   &   $   [[2,0]_2,2]_{60}    $   &   $\xeleven$  \\
 301--309    &   40.91   &   $   [[2,1]_j,\Lambda]_{40}$ (mixed)   &   $\xtwelve$  &   &   115--120    &   33.71   &   $   [[2,0]_2,3]_{60}    $   &   $\xten$ \\
 310--318    &   41.03   &   $   [[2,1]_j,\Lambda]_{40}$ (mixed)   &   $\xfifteen$ &   &   121--132    &   35.18   &   $   [[2,1]_j,\Lambda]_{60}$ (mixed)   &   $\xsixteen$ \\
 319--327    &   41.07   &   $   [[2,1]_j,\Lambda]_{40}$ (mixed)   &   $\xtwelve$  &   &   133--138    &   35.22   &   $   [[2,0]_2,4]_{60}    $   &   $\xeleven$  \\
  328--336    &   41.45   &   $   [[2,1]_j,\Lambda]_{40}$ (mixed)   &   $\xtwelve$  &   &   139--150    &   36.70   &   $   [[2,1]_j,\Lambda]_{60}$ (mixed)   &   $\xsixteen$ \\
 337--345    &   41.54   &   $   [[2,1]_j,\Lambda]_{40}$ (mixed)   &   $\xtwelve$  &   &   151--154    &   36.75   &   $   [[2,0]_2,L]_{60}\ (L=4,6)$   &   $\xnine$    \\
 346--354    &   41.72   &   $   [[2,1]_j,\Lambda]_{40}$ (mixed)   &   $\xfifteen$ &   &   155 &   37.17   &   $   [[0,0]_0,6]_{60}    $   &   $\xone$ \\
 355--363    &   41.84   &   $   [[2,1]_j,\Lambda]_{40}$ (mixed)   &   $\xtwelve$  &   &   156--161    &   37.21   &   $   [[2,0]_2,5]_{60}    $   &   $\xten$ \\
     &       &   $       $   &       &   &   162--173    &   37.78   &   $   [[2,1]_j,\Lambda]_{60}$ (mixed)   &   $\xsixteen$ \\
     &       &   $       $   &       &   &   174--179    &   38.72   &   $   [[2,0]_2,6]_{60}    $   &   $\xeleven$  \\
\hline\\[-0.35cm]
\hline
\end{tabular}
}
\end{table}

%\bibliography{mybib}

\end{document}

%% file: metmet_text.tex
%-------------------------------------------------------------------
\section{Introduction}
\noindent
Detailed understanding of hydrocarbon interactions has been a challenging subject for physical chemistry.
The alkyl-alkyl interaction is common in molecular systems since alkyl chains are ubiquitous in organic, bio-, and materials chemistry. 
In atmospheric and astrochemical processes the smallest hydrocarbon, methane, has utmost importance.

For a microscopic characterization of molecular interactions, including pair interactions,  high-resolution spectroscopy\cite{HRSBook} of molecular complexes is one of the powerful approaches. High-resolution spectroscopy provides detailed information on the energy level structure, which can be quantitatively analyzed with respect to quantum nuclear motion computations using the interaction potential energy surface.\cite{Te16jcp,Ca16jcp,MaAvDa23}

Electronic structure theory has been used to compute the \emph{ab initio} intermolecular potential energy surface (PES)\cite{MePiSz16} for a variety of complexes, and the local minima of the PES define the equilibrium structures.\cite{StructBook10}
At the same time, the intermolecular PES of alkyl-alkyl systems is flat, corresponding to small, attractive forces (beyond the van-der-Waals radius), and hence the dynamical properties of hydrocarbon systems are dominated by large-amplitude motions of the atomic nuclei, which is correctly accounted for by quantum mechanics.
In particular, the zero-point vibrational energy of the light hydrocarbon systems is comparable to or often exceeds the `small barrier heights’ of the flat interaction PES. As a result, the zero-point state is delocalized over several shallow PES wells, and the simple concept of a rigid molecular structure with small amplitude vibrations (usually treated as a perturbation) is not applicable.

The methane dimer is the smallest hydrocarbon aggregate to feature alkyl-alkyl interactions. An \emph{ab initio} potential energy surface has been recently reported in Ref.~\citenum{dimers} in relation to all three possible molecular dimers that can be formed by the methane and the water molecules, with relevance to methane clathrates.

Regarding the methane dimer, an infrared spectroscopy study was previously initiated \cite{haPhD05}, but the spectral peaks corresponding to internal rotational states have not been assigned/analyzed in detail, probably due to the complicated predissociative quantum dynamics. 

Recent progress with advanced imagining techniques made it possible to record the rotational Raman spectrum of (apolar) hydrocarbon complexes,\cite{MiSaToOh22,OhToMi22,MiTaKaToOh23} including the rotational Raman spectrum of the ethane dimer and trimer.  

In this paper, we report the computed rotational Raman spectrum of the five lowest-energy spin isomers of the methane dimer based on variational rovibrational computations on an \emph{ab initio} intermolecular potential energy surface\cite{dimers} and polarizability transition moments using a simple polarizability model.

The computational methodology is summarized in Sec.~\ref{sec:rovib}. 
The molecular symmetry (MS) group and the spin statistical weights are discussed in Sec.~\ref{sec:msgroup}.
Sec.~\ref{sec:anal} is about the analysis and assignment of the computed rovibrational wave functions. 
The paper ends (Sec.~\ref{sec:predlist}) with a short analysis of the predicted polarizability transitions, with a comprehensive list deposited as Supplementary Information to facilitate future experimental work.

%\newpage
%-------------------------------------------------------------------

\section{Theoretical and computational details}

\subsection{Numerical solution of the intermolecular rovibrational Schrödinger equation \label{sec:rovib}}

The intermolecular, rovibrational Schrödinger equation of \metmet\ has been solved using the GENIUSH program \cite{MaCzCs09,FaMaCs11}. The program already has several applications  
for semi-rigid and floppy molecular systems \cite{FaCsCz13,14FaMaCs,SaCsAlWaMa16,ArNOp,FaQuCs17,SaCs16,SaCsMa17,dimers,FeMa19,AvMa19,AvMa19b,AvPaCzMa20,DaAvMa21mw,fad21},
recently reviewed in Ref.~\citenum{MaAvDa23}. 
The Podolsky form of the rovibrational Hamiltonian was used during the computations,   
\begin{equation}
   \label{eq:hamil}
   \begin{split}
    & \hat{H} 
    = 
    \frac{1}{2} \sum_{k=1}^D \sum_{l=1}^D \tilde{{g}}^{-1/4} \hat{p}_k G_{kl}\tilde{{g}}^{1/2}  \hat{p}_l\tilde{{g}}^{-1/4} \\
    & + \frac{1}{2} \sum_{k=1}^D \sum_{a=1}^3(\tilde{{g}}^{-1/4} \hat{p}_k G_{k,D+a}\tilde{{g}}^{1/4} + \tilde{{g}}^{1/4} G_{k,D+a}\hat{p}_k \tilde{{g}}^{-1/4})\hat{J}_a \\
    & + \frac{1}{2} \sum_{a=1}^3 G_{D+a,D+a}\hat{J}_a^2\\
    & + \frac{1}{2} \sum_{a=1}^3 \sum_{b>a}^3 G_{D+a,D+b}[\hat{J}_a,\hat{J}_b]_+ + \hat{V} \; ,
    \end{split}
\end{equation}
where  $\hat{J}_a$ is the $a=1(x),2(y),3(z)$ component of the body-fixed total angular momentum operator,  $\hat{p}_k=-\text{i}\partial/ \partial q_k$ is defined for every active $q_k\ (k=1,2,\ldots,D)$ vibrational coordinate, and $V$ labels the potential energy surface. 

The user-defined curvilinear coordinates and the body-fixed frame are encoded in the rovibrational $\bos{g}\in\mathbb{R}^{(D+3)\times(D+3)}$ 
mass-weighted metric tensor and $D\leq 3N-6$ is the number of active vibrational degrees of freedom of the $N$-atomic system.
The kinetic energy operator coefficients, $G_{kl}=(\bos{g}^{-1})_{kl}$ and $\tilde{{g}}=\text{det}(\bos{g})$ are obtained from direct computation of $\bos{g}$ over a grid, 
\begin{equation}
  g_{kl}= \sum_{i=1}^N m_i \bos{t}^\text{T}_{ik} \bos{t}_{il}, 
   \quad\quad\quad k,l = 1,2,...,D+3 
  \label{eq:gmxt}
\end{equation}
with the vibrational and rotational $\bos{t}$-vectors,
\begin{equation}
\bos{t}_{ik} = \frac{\partial \bos{r}_i}{\partial q_k}, 
   \quad\quad\quad k = 1,2,...,D
   \label{eq:tvib}
\end{equation}
\begin{equation}
\bos{t}_{i,D+a} = \bos{e}_a \times \bos{r}_i, 
   \quad\quad\quad a = 1(x),2(y),3(z) \; ,
   \label{eq:trot}   
\end{equation}
respectively, where $\bos{r}_i$ are the body-fixed Cartesian coordinates for the $i$th atom and $\bos{e}_a$ labels the unit vector in the body-fixed frame.

%%%%%%%%%%%%%%%%%%%%%%%%%%%%%%%%%%%%%%%%%%%%%%%%%%%%%%%%%%%%%%%%%%%%%%%%%%%%%%%%%%%%%%%
\subsection{Definition of the vibrational coordinates and grid representation \label{sec:coord}}

\begin{figure}
  \includegraphics[width=8cm]{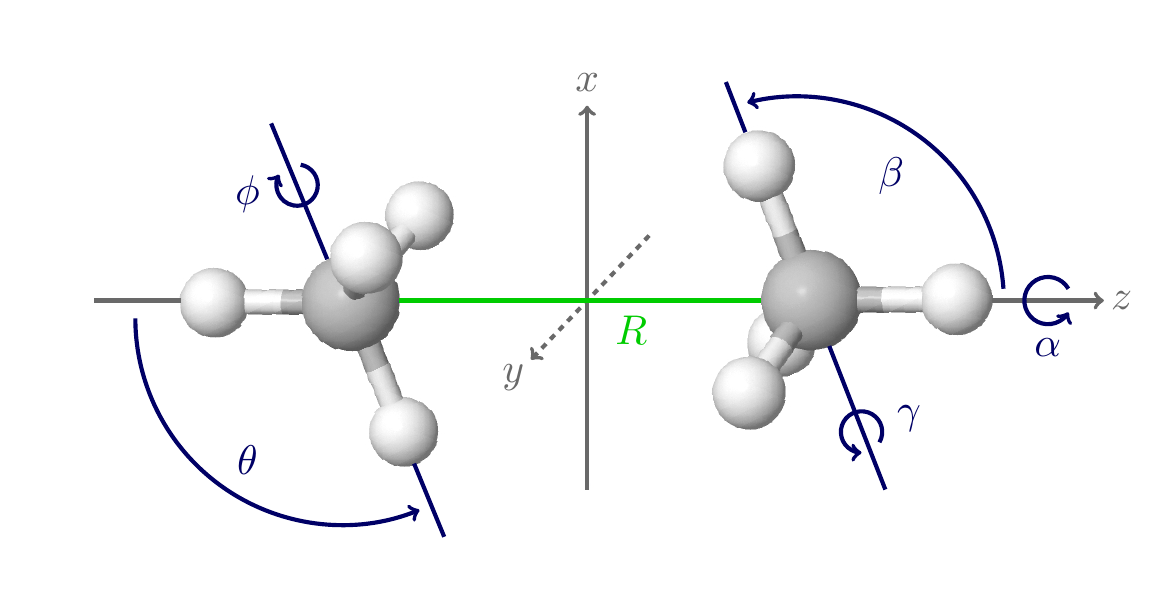}
  \caption{%
    Intermolecular coordinates, $(R,\theta,\phi,\alpha,\beta,\gamma)$, used to describe the intermolecular vibrational quantum dynamics of the \metmet\ dimer shown for the example of the global minimum (GM) structure of the MM19-PES \cite{dimers}.
    $R\in[0,\infty)$, $\phi,\alpha,\gamma\in[0,2\pi)$, $\theta,\beta\in[0,\pi]$.
    The structure of the methane fragments is fixed during the computations.
    \label{fig:coord}
  }
\end{figure}

\metmet\ has $N=10$ nuclei with a total number of 24 vibrational degrees of freedom. 
To study its intermolecular rotational-vibrational dynamics, a 6-dimensional (6D) model was defined by fixing the methane monomer structures (\emph{intra}molecular degrees of freedom). 
In this model, the active \emph{inter}molecular coordinates, Figure~\ref{fig:coord}, describe the relative position and orientation of the two fragments and are defined by
(a) the  distance between the centres of mass of the two methane molecules, $R \in [0,\infty)$; 
(b) two angles to describe the orientation of monomer one, $\cos{\theta} \in [-1,1]$ and $\phi \in [0,2\pi)$ (out of the three Euler angles, the first Euler angle is fixed at 0); 
and 
(c) three Euler angles describing the orientation of monomer two, $\alpha \in [0,2\pi)$, $\cos{\beta} \in [-1,1]$ and $\gamma \in [0,2\pi)$.

Regarding the frozen structures of the two CH$_4$ molecules, we used in the vibrationally-averaged effective values that had also been used during the MM19-PES development \cite{dimers}. The effective carbon-proton distance was $r(\text{C--H}) = 1.110\ 020\ 996\ \text{\r{A}}$, and 
$\cos\alpha(\text{H--C--H}) = -1/3$ corresponding to a regular tetrahedron.
In all computations, we used the atomic masses, $m(\text{H}) = 1.007825\ \text{m}_\text{u}$ and $m(\text{C}) = 12\ \text{m}_\text{u}$. 

The matrix representation of the Hamiltonian, Eq.~\eqref{eq:hamil}, is constructed using the discrete variable representation (DVR) scheme,\cite{LiCa00} 
and an iterative Lanczos eigensolver was employed \cite{MaSiCs09} to converge the lowest (few hundred) eigenstates of the Hamiltonian matrix.

For our coordinate choice (Fig.~\ref{fig:coord}), the kinetic energy operator (KEO) is singular at $\cos\theta = \pm1$ and $\cos\beta = \pm1$. We have extensively studied the effect of these types of singularities on the energy levels convergence rate in the case of the \metwat\ dimer in Ref.~\cite{DaAvMa21mw}. The cot-DVR approach \cite{ScMa10} (including also two sine functions in the basis set) was found to be computationally efficient for converging the singular degrees of freedom. The cot-DVR representation was originally developed by Schiffel and Manthe,\cite{ScMa10} and its first molecular application (H$_2$O and Ar$\cdot$HF) was reported by Wang and Carrington.\cite{WaCa12}

The convergence of the computed rovibrational energy levels with respect to the DVR points has been tested using different reduced-dimensionality models, from 1D to 6D. 
The `optimal' grid (and basis) parameters are collected in Table~\ref{tab:metmet_dvr}.
This grid includes $1.8\cdot 10^6$ points, which equals the number of basis functions in DVR.
We have tested the convergence provided by this grid and basis size for the $J=0$ rotational quantum number. During the course of the convergence tests, the largest grid size included $7.6\cdot 10^6$ points and provided energy levels (in the studied energy range) converged better than 0.01~\cm.
With respect to this large computation, the convergence error of the absolute energies obtained with the `optimal' basis (used for $J\geq 0)$ is 0.1~\cm\ (Table~\ref{tab:metmet_dvr}), while the energies relative to the zero-point vibrational energy (ZPVE) are converged better than 0.01~\cm. 
Since we were interested in rotational transitions up to $J=6$ rotational quantum number,
and the dimensionality of the Hamiltonian matrix increases by a factor of $\sim(2J+1)$,
we used the `optimal' grid (Table~\ref{tab:metmet_dvr}) in the rovibrational computations. 
Using this `optimal' grid and basis, it was possible to compute the 400 lowest-energy states for every $J$ value up to  $J=6$ within a reasonable amount of computer time (within a few weeks on multiprocessor CPUs).

\begin{table}%[h]
\caption{
    The coordinate intervals and grid representations used in the GENIUSH intermolecular rovibrational computations for the methane dimer.
    \label{tab:metmet_dvr}
    }
%\singlespacing
    \centering
      \raisebox{0cm}{%
      \begin{tabular}{@{}l r@{\ \ \ \ } c@{\ \ \ \ } cc@{\ \ \ \ } c c@{}}
\hline\\[-0.35cm] 
%\hline\\[-0.15cm] 
  \raisebox{-0.40cm}{Coord.} & 
  \raisebox{-0.40cm}{Eq.$^\text{a}$} & 
  \multicolumn{3}{c}{DVR} & & 
  \raisebox{-0.4cm}{No. points}  \\[-0.35cm]
\cline{3-5}\\[-0.35cm]
           &                   & Type &  & Interval &  \\%[0.2cm]
\hline\\[-0.35cm]
$R\ [\textup{\r{A}}]$ & 3.638 & PO-Laguerre$^\text{b}$&  & [2.5,7.0] && 7 \\
$\theta\ [^{\circ}]$  & 109.471 & cot-DVR$^\text{c}$ &  & (0,180) && 13 \\
$\phi\ [^{\circ}]$   & 180.000 & Fourier &  & [0,360) && 13 \\
$\alpha\ [^{\circ}]$ & 120.000 & Fourier &  & [0,360) && 9 \\
$\beta\ [^{\circ}] $   & 109.471 & cot-DVR$^\text{c}$ &  & (0,180) && 13 \\
$\gamma\ [^{\circ}]$ & 180.000 & Fourier &  & [0,360) && 13 \\%[0.15cm]
%\hline\\[-0.2cm] 
\hline
\end{tabular}
      }
\begin{flushleft}
$^\text{a}$~Equilibrium structure of the MM19-PES \cite{dimers} in internal coordinates. The equilibrium rotational constants are $A=2.545\ 37$~\cm\ and $B=C=0.149\ 404$~\cm.\\
$^\text{b}$~Potential-optimized DVR using 300 primitive grid points.\\
$^\text{c}$~The cot-DVR was constructed with two sine functions.\\
\end{flushleft}
\end{table}

\subsection{Molecular symmetry group and spin statistics \label{sec:msgroup}}

The molecular symmetry (MS) group of \metmet\ is $G_{576}$. The corresponding character table (deposited in the Supplementary Information) has been generated using the GAP program~\cite{GAP} following the instructions of Ref.~\citenum{ScLe04}. 
We note that the ordering of the rows and columns of this character table is different from Ref.~\citenum{OdAlCuDy81}.
The spin statistical weights taken from Ref.~\citenum{ScLe04} are also listed in the same table.   
The symmetry labels of the computed rovibrational wave functions are assigned using the coupled-rotor decomposition scheme
and the symmetry assignment of the methane-methane CR functions is carried out by generalizing the procedures of Ref.~\citenum{SaCsMa17,FeMa19}.

The (proton) spin states of a single methane molecule correspond to a total proton spin quantum number $I=0$ (meta), $I=1$ (ortho), and $I=2$ (para) \cite{BuJe98}.
The lowest-energy spatial functions, which correspond to these spin states by satisfying the generalized Pauli principle,  are the $j=0$ (meta, $I=0$), 
$j=1$ (ortho, $I=1$), and $j=2$ (para, $I=2$) rotational states of methane.\cite{BuJe98,hougen01}
Since the monomer (proton) spin is a good quantum label of the dimer, we can distinguish 
meta-meta (m-m), meta-ortho (m-o), meta-para (m-p), ortho-ortho (o-o), ortho-para (o-p), and para-para (p-p) spin isomers of the complex.

\section{Analysis of the rovibrational wave function \label{sec:anal}}

Different strategies have been used to analyze the computed rovibrational wave functions of \metmet, including 
%computation of expectation values \ref{sec:rexp}, inspection of wave function cuts and nodal features, rigid-rotor decomposition---vibrational parent assignment \ref{sec:rrd}, and rotational parent assignment with (approximate) $K$ labels \ref{sec:klabels}, coupled-rotor decomposition, and MS group assignment \ref{sec:crd}.
%
    computation of expectation values (Sec.~\ref{sec:rexp});
    inspection of wave function cuts and nodal features (Sec.~\ref{sec:cuts});
    rigid-rotor decomposition---vibrational parent assignment (Sec.~\ref{sec:rrd});
    rotational parent assignment with (approximate) $K$ labels (Sec.~\ref{sec:klabels});
    and coupled-rotor decomposition and MS group assignment (Sec.~\ref{sec:crd}).

\subsection{Intermolecular expectation values \label{sec:rexp}}

\begin{figure}%[h]
\centering
  \includegraphics[width=0.9\linewidth]{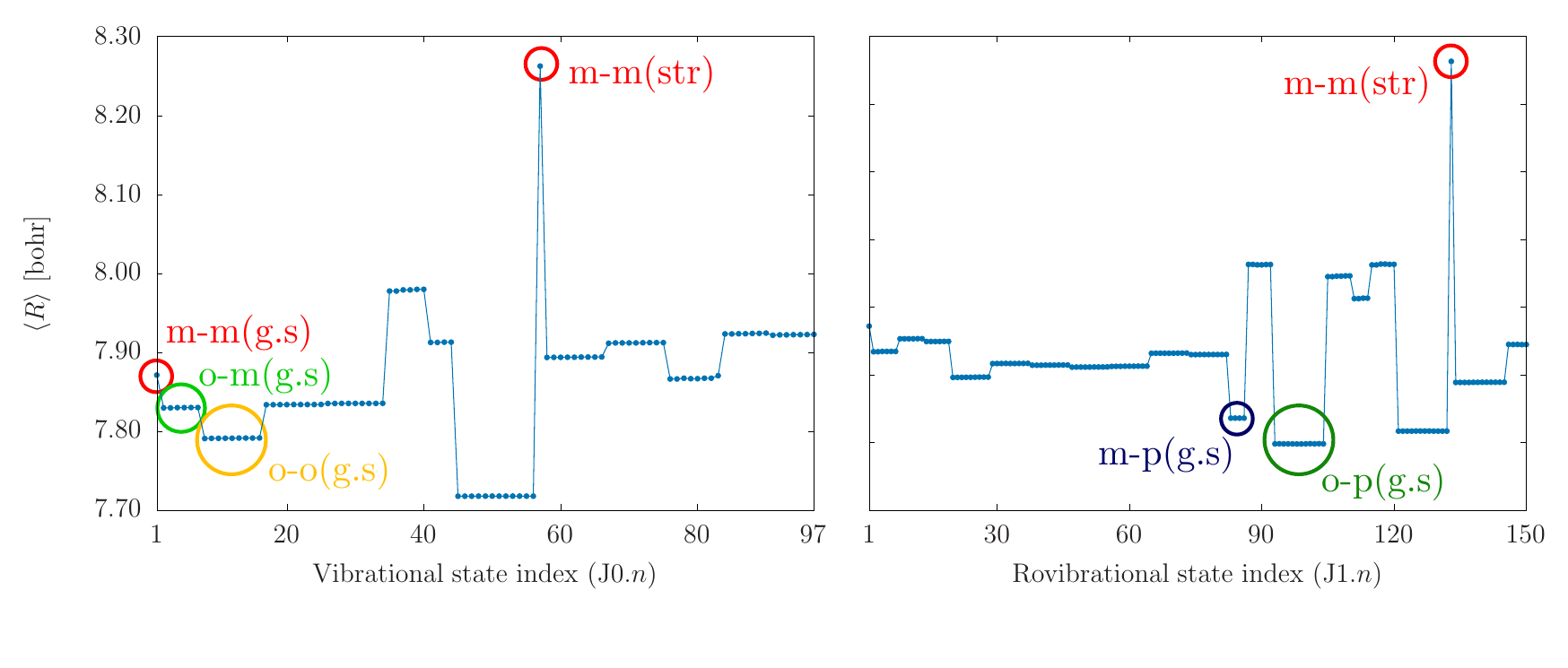}
  \caption{
    The expectation value of the distance of the carbon atoms in the methane fragments.
  \label{fig:rexp}}
\end{figure}

We have computed the expectation value of the $R$ intermolecular distance between the two methane moieties. 
In Fig.~\ref{fig:rexp}, the value of the expectation value, $\langle R \rangle$, is shown for a few  of dozens of states for $J=0$ and $J=1$, highlighting the ground states (g.s) for each spin isomers
(based on assignment in Sec.~\ref{sec:crd}).
In Fig.~\ref{fig:rexp}, the stretching (str) state can also be clearly identified by the increased value of the average intermolecular distance (that is further confirmed by identifying the node along the $R$ degree of freedom in an appropriate wave function cut).
\subsection{Wave function cuts and node counting \label{sec:cuts}}
\begin{figure}%[H]
    \centering
    \includegraphics[width=\linewidth]{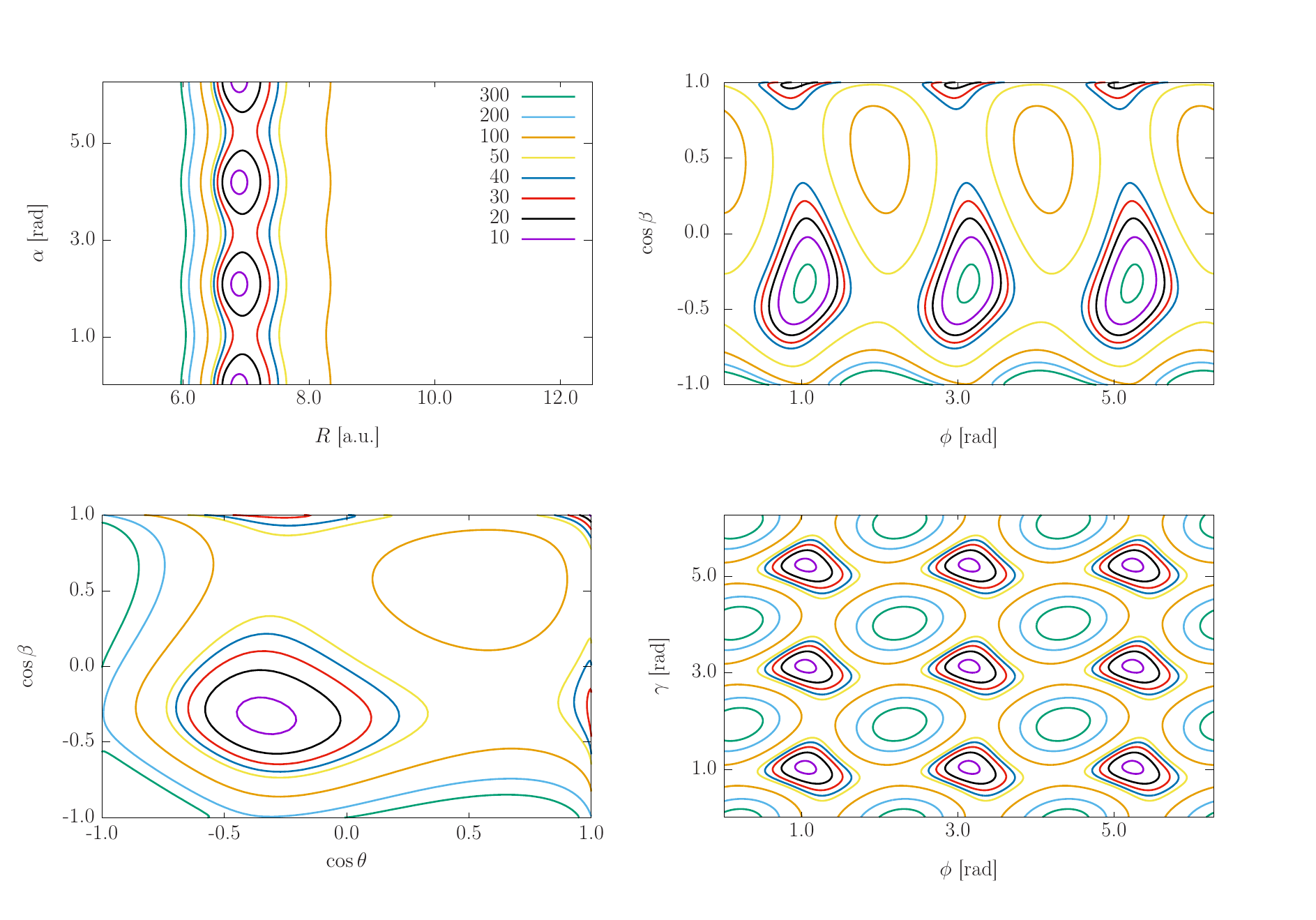} %\\[-0.75cm]
    \caption{Selected 2-dimensional cuts of the MM19-PES~\cite{dimers} along the intermolecular coordinates of \metmet.
    The rest of the coordinates are fixed at their equilibrium value (Table~\ref{tab:metmet_dvr}).
    The energy scale, in \cm, for the contour line is shown in the legend of the upper-left subfigure ($R-\alpha$ cut).
    }
    \label{fig:2dpes}
\end{figure}

\begin{figure}%[H]
    \centering
    \includegraphics[width=8cm]{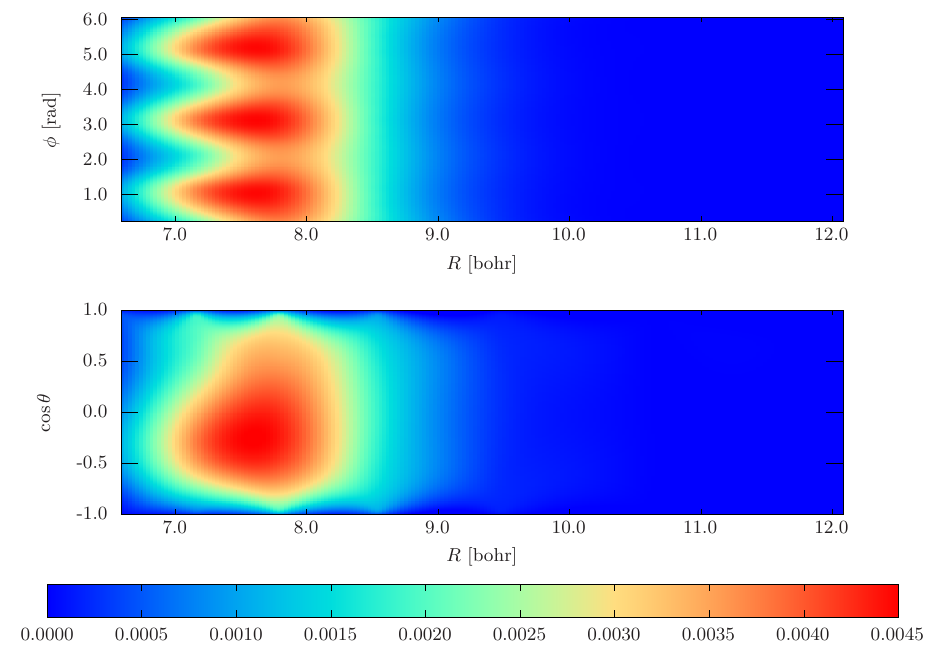}
    \caption{%
    2D cut of the wave function of the lowest energy state of the \metmet\ meta-meta spin isomer, $[0,0]$, which corresponds to the ground vibrational state,
    J0.1.
    Wave function cuts for the other four spin isomers are deposited as Supplementary Information.
    } 
    \label{fig:mm_wf}
\end{figure}

The anisotropy of the interaction can be visualized by selected 2D cuts of the MM19-PES~\cite{dimers} (Fig.~\ref{fig:2dpes}).
The effect of this anisotropy in the wave functions has been studied for the lowest-energy states of each spin isomer.
Two wave function cuts are shown in Figs.~\ref{fig:mm_wf} for the ground state of the meta-meta spin isomer, 
while the equivalent cuts for the other spin isomers are included in the Supplementary Information.
%If the state corresponds to a degenerate block, only one state is plotted.

%\clearpage
    
%

%\subsection{Rigid rotor limit and vibrational parent, rotational parent, and assignment of the $K$ labels } 

\subsection{Vibrational parent analysis, rigid rotor limit \label{sec:rrd}}

The first limiting model that we consider---which is an inefficient model for this weakly-bound system---is based on the rigid rotor model of molecular rotations. 
The rigid rotor decomposition (RRD) \cite{NMD10} scheme is based on the evaluation of the overlap of the $m$th rovibrational wave function, $\Psi_m^{J>0}$, and the product of vibrational functions, $\psi_n^{J=0}$, and rigid-rotor functions, $\phi_l^{{\rm RR},J}$, as
\begin{align}
%\label{eq:rrd}
 S_{nl,m} = \bra{\psi_n^{J=0} \cdot \phi_l^{{\rm RR},J}} \ket{\Psi_m^{J>0}} 
 \label{eq:rrd1}
\end{align}
with
\begin{align}
\label{eq:rrd2}
 \ket{\psi_n^{J=0} \, . \, \phi_l^{{\rm RR},J}} =
    \ket{\psi_n^{J=0}} \otimes \phi_l^{\text{RR},J} ,
\end{align}
where the $\phi_l^{{\rm RR},J}$ rigid-rotor functions are the Wang functions \cite{Za98}.
It is necessary to note that the RRD overlaps, Eq.~(\ref{eq:rrd1}), depend on the body-fixed frame that was defined in Sec.~\ref{sec:coord}. 

\begin{figure}%[H]
    \centering
    \includegraphics[width=9cm]{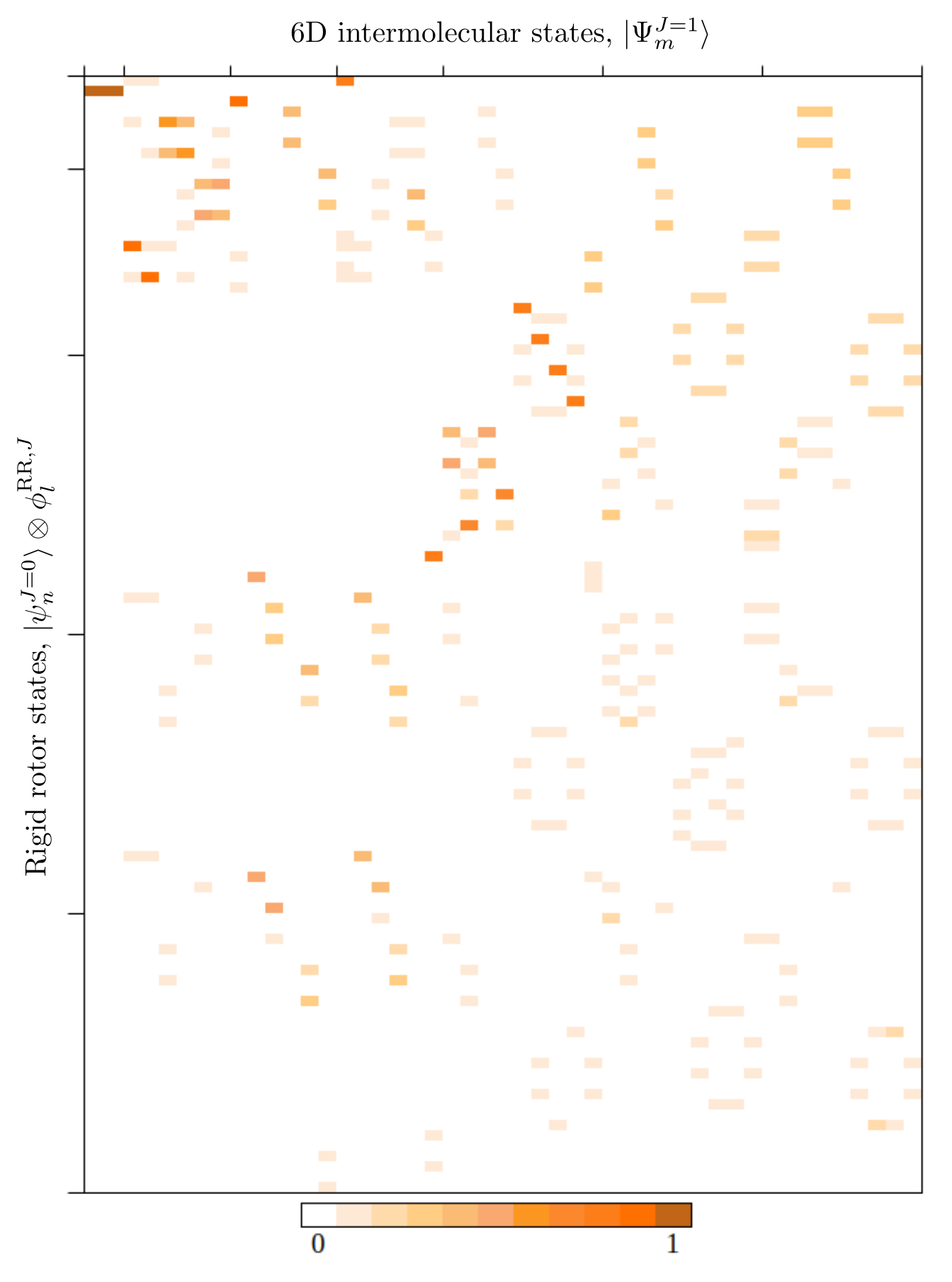}
    \caption{%
    RRD overlap coefficients computed for the $J=1$ states of \metmet. 
    The full-dimensional (intermolecular, 6D) rovibrational states, $J=1$, are in columns.
    }
    \label{fig:rrd}
\end{figure}

If there is a single (or few dominant) RRD coefficient, $|S_{nl,m}|^2$, for a rovibrational state,
then the vibrational `parent'(s) of the state can be unambiguously identified. 
For the methane dimer the RRD matrix, Fig.~\ref{fig:rrd}, is relatively `pale', which highlights that the rigid rotor model is not a good model for this system.

Although a dominant $\psi_n^{J=0}$ vibrational state cannot be assigned to a rovibrational state, we can still aim for the identification of the rotational function, $\phi_l^{\text{RR},J}$, which gives dominant contribution to the rovibrational state. 
The GENIUSH program uses Wang functions \cite{Wa29} (for a modern introduction, see for example \cite{BuJe98}) as rotational basis functions, because the Hamiltonian matrix is real in this representation. 
The Wang functions \cite{BuJe98} are defined as linear combination of the orthogonal symmetric top (rigid rotor) functions, 
\begin{align}
\theta^{(JM)}_{K\tau}
%\ket{JK\tau M}
=
\left\lbrace 
\begin{array}{@{}cc@{}}
    & \frac{1}{\sqrt{2}} \left[ \ket{J\bar{K}M} + \ket{J-\bar{K}M} \right], \quad \text{for even}~\bar{K}, \quad \tau=0  \\
    & \frac{\iim}{\sqrt{2}} \left[ \ket{J\bar{K}M} - \ket{J-\bar{K}M} \right], \quad \text{for even}~\bar{K}, \quad \tau=1 \\
    & \frac{1}{\sqrt{2}} \left[ \ket{J\bar{K}M} - \ket{J-\bar{K}M} \right], \quad \text{for odd}~\bar{K},  \quad ~\tau=0\\
    & \frac{\iim}{\sqrt{2}} \left[ \ket{J\bar{K}M} + \ket{J-\bar{K}M} \right], \quad \text{for odd}~\bar{K},  \quad ~\tau=1\\
\end{array}
\right. \; ,
\end{align}
and $\Theta_{00}^{(JM)}=\ket{J0M}$ ($\Theta_{01}^{(JM)}=0$).
%and $\ket{J00M}=\ket{J0M}$ ($\Theta_{01}^{(JM)}=0$).

In practice, this means that the rovibrational basis set has $(2J+1)$ vibrational `sub-blocks', and each sub-block is characterized by a value of $K$ ($\bar{K}$ absolute value) and $\tau$.

\subsection{Rotational parent analysis, $K$ label assignment \label{sec:klabels}}

Although the methane dimer is a symmetric top at equilibrium structure, the $K$ label is only an approximate quantum number due to the highly fluxional character of the system. 

We assigned the $K$ (and $\tau$) labels of the Wang functions to the intermolecular rovibrational states computed in this work, by calculating the sum of the absolute value squared coefficients over the different vibrational `sub-blocks' (corresponding to different Wang functions) of the rovibrational wave function.

So, the contribution of the 
$\Theta_{K\tau}^{(JM)}$
%$|JK\tau M\rangle$ 
Wang function to the $n$th rovibrational wave function with $J$ rotational angular momentum quantum number, $\Psi^{(J)}_n$ (the energy is independent of $M$) is measured by the quantity
\begin{align}
  \tilde\kappa^{(J)}_K(\tau) 
  &= 
  \sum_{i=1}^{N_\mathrm{grid}^\mathrm{vib}} 
    | \psi^{(J)}_{n,(K,\tau)} (\xi_i) |^2 \; 
  \label{eq:kappaKtau}
\end{align}
and we can sum for the $\tau=0,1$ values to have a measure only for the $K$ label,
\begin{align}
  \kappa^{(J)}_K
  &=
  \sum_{\tau=0,1} 
    \tilde\kappa^{(J)}_K(\tau) \; .
  \label{eq:kappaK}
\end{align}
In Eqs.~(\ref{eq:kappaKtau})--(\ref{eq:kappaK}), $\psi^{(J)}_{n,(K,\tau)} (\xi_i)$ labels the $K\tau$ sub-block of the rovibrational eigenvector corresponding to the $\xi_i$ grid point in the multi-dimensional DVR grid,  $i=1,\ldots,N_\text{grid}^\text{vib}$ (Table~\ref{tab:metmet_dvr}).

\begin{figure}%[h]
  \includegraphics[width=9cm]{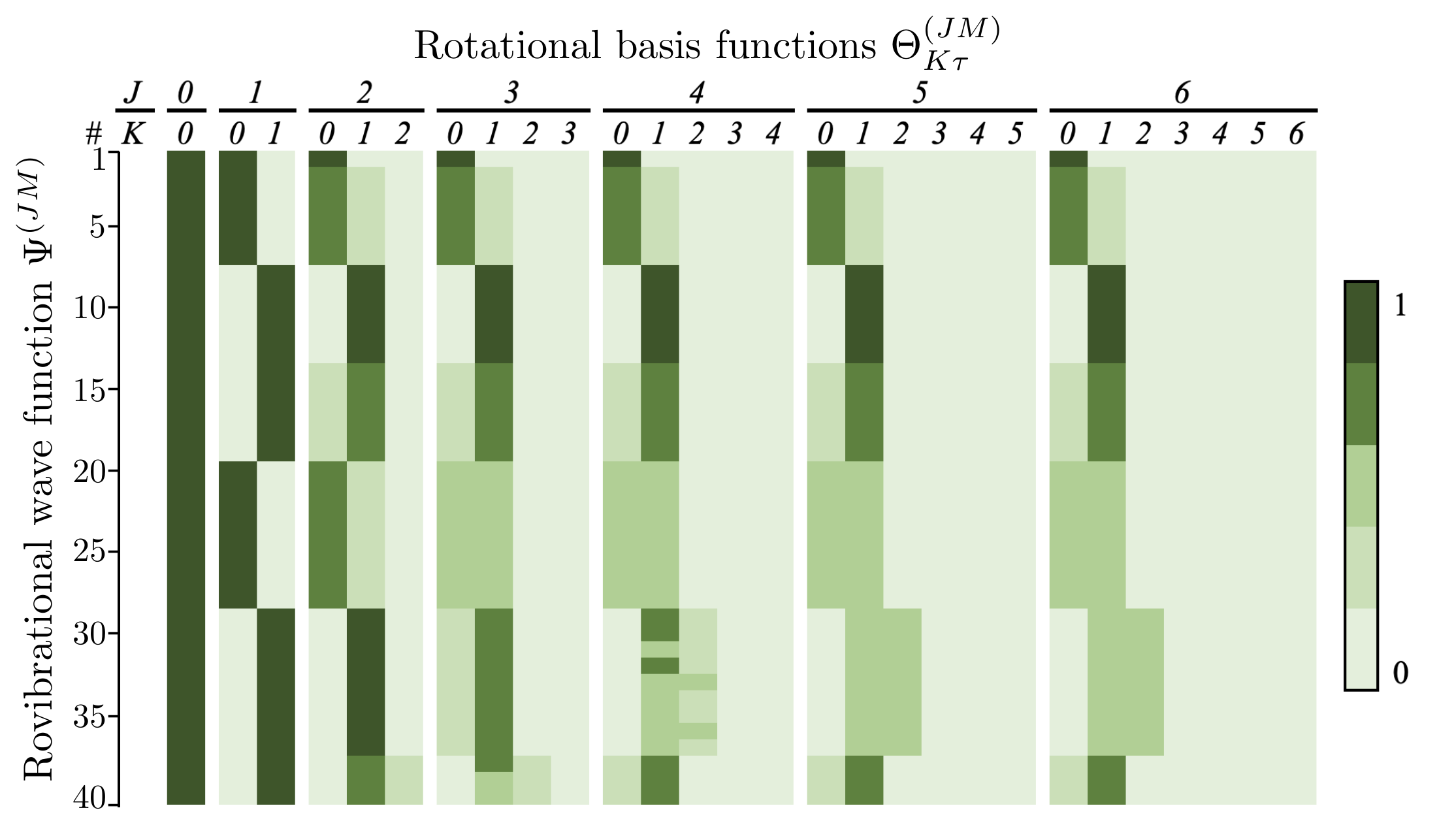}
  \caption{%
    Although the equilibrium structure of (CH$_4$)$_2$ is a symmetric top, the assignment of $K$ labels to the rovibrational states becomes ambiguous already for low-energy $J=3$ rovibrational states of this floppy complex.
    \label{fig:Kassign}
  }
\end{figure}

Based on this simple calculation, schematized in Fig.~(\ref{fig:Kassign}), it was possible to unambiguously assign a $K$ value for several rovibrational states, but beyond $J>2$, there are (low-energy) states for which the assignment is ambiguous.

%~\\[1cm]

\subsection{Coupled rotor limit \label{sec:crd}}
Another limiting model used to characterize the rovibrational dynamics of \metmet\ is the coupled-rotor (CR) limit.
The coupled-rotor decomposition (CRD) scheme \cite{SaCsMa17} defined based on this limiting model is used to assign monomer (methane) rotational states to the complex of two methane moieties. 

The CRD is based on measuring the similarity (as a special overlap of the wave functions) between the 6D intermolecular rovibrational wave functions and the 5D angular functions of free-rotating monomers (without interaction) fixed at a given distance. As a result of a second, 5D computation, the CR functions become available in exactly the same (DVR grid) representation as the (angular part of the) 6D intermolecular states, and hence, their overlap can be straightforwardly computed. 

The MS group assignment of the dimer states is carried out based on the symmetry of the CR functions.

The CR states (5D) are characterized by two non-interacting rigid rotors fixed at some distance (at the equilibrium distance or at a very large distance), where the angular momenta of the two sub-systems are coupled between themselves and with the angular momentum of the effective diatom rotation.

The CR states (computed at some finite fixed intermolecular distance, $R$) are assigned based on their energies and using the monomer rotational energies and the energy correction due to the rotation of the effective diatom connecting the methane centres of masses.\cite{SaCsMa17,BrAvSuTe83} The diatom term vanishes if the $R$ fixed monomer separation is very large (in practice, $R\geq 100$~bohr).
As a result, the CR states are assigned with the monomer rotational angular momentum quantum numbers, 
$j_1$ and $j_2$. The monomers' angular momenta are coupled to an internal rotational angular momentum with quantum number $j$. This internal angular momentum is coupled with the rotational angular momentum of the effective diatom (corresponding to the relative rotation of the two methane fragments), with the rotational quantum number $\Lambda$, to the total rotational angular momentum of the complex, with the total rotational angular momentum quantum number, $J$, and its projection quantum number, $M$. This angular momentum coupling scheme is labelled as
\begin{align}
 \left[ 
 \left[ 
    j_{1}, j_{2} 
 \right]_j ,
 \Lambda
 \right]_{JM} \; ,
\end{align}
which is finally used to label the CR states to characterize their angular dependence.

The overlap between the 5D CR states and the 6D intermolecular states is computed for every $J$ value as
\begin{equation}
  \text{CRD}^{(J)}_{nm} 
  = 
  \sum_{r=1}^{N_R} \Bigg| 
    \sum_{k=-J}^J 
    \sum_{o =1}^{N_\Omega}\tilde{\Psi}^{(J)}_{m,k}
    (\rho_r,\omega_o) \, \cdot \, \tilde{\varphi}^{(J)}_{n,k}(\omega_o) \Bigg|^2  \; ,
  \label{eq:crd}
\end{equation}
where $\bar{\Psi}^{(J)}_m$ is the $m$th rovibrational state, depending 
on the intermolecular distance $R$ 
with grid points $\rho_r\ (r=1,\ldots,N_R)$ and 
on the five (cos) angles $\Omega=\cos\theta,\phi,\alpha,\cos\beta,\gamma$ 
with grid points $\omega_o\ (o=1,\ldots,N_\Omega)$, and $\bar{\varphi}^{(J)}_n$ is the $n$th CR function depending only on the angular part over the same (angular) DVR grid points.

\begin{figure}%[H]
    \centering
    \includegraphics[width=9cm]{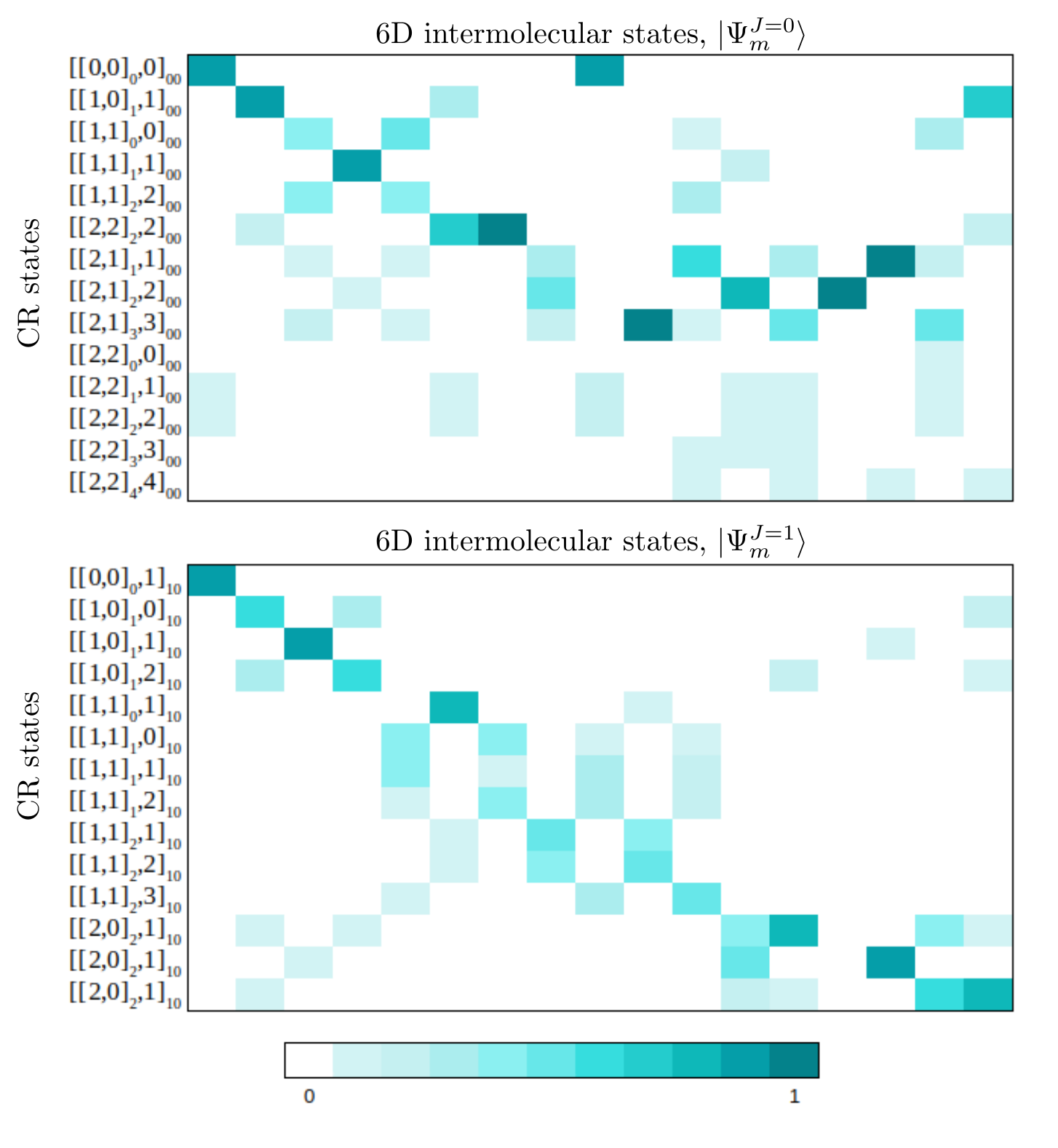}
    \caption{%
    CRD overlap coefficients computed for the $J=0$ and $J=1$ states of the \metmet\ dimer. The coupled rotor functions are in rows, and the full-dimensional (intermolecular, 6D) states are in columns.
    }
    \label{fig:mm_crd}
\end{figure}

The CRD matrices have two key properties: (a) the sum of the elements in a column is 1, if a large number of (infinitely many) CR functions is used, \emph{i.e.}, for each $\bar{\Psi}^{(J)}_m$ state, the sum of the CRD contributions over all CR states is 1; and (b) the sum of the elements in a row can be larger than 1, \emph{i.e.}, one CR function can contribute (and even be dominant) in several $\bar{\Psi}^{(J)}_m$ states. Figure~\ref{fig:mm_crd} vizualizes the CR overlap matrix elements, Eq.~(\ref{eq:crd}), for $J=0$ and $J=1$.

Based on these overlap matrices, we assigned CR labels to all 6D intermolecular rovibrational states. 
The transformation properties of the coupled-rotor functions (5D basis function without intermolecular interaction) under the permutation-inversion operators of the molecular symmetry group can be derived 
based on formal arguments (we generalized the calculation of Refs.~\citenum{SaCsMa17} and \citenum{FeMa19} carried out for the CH$_4\cdot$H$_2$O and CH$_4\cdot$Ar complexes). 
Then, by using the CRD assignment of the 6D states \emph{(e.g.,} Fig.~\ref{fig:mm_crd}), we attach an irrep label to every 6D state.

Regarding the formal symmetry analysis of the CR functions, 
it is practical to distinguish between states in which the two monomers are in the same rotational state, $j_1=j_2$; and states which have different monomer rotational quantum numbers, $j_1\neq j_2$. 
We consider the set of the $[[j_1,j_2]_j,\Lambda]_{JM}$ CR functions with $k_1=-j_1,\ldots,j_1$ and $k_2=-j_2,\ldots,j_2$ values (in $|j_i,k_i,m_i\rangle$) as a representation of the MS group. 
If $j_1\neq j_2$, then we include both $[[j_1,j_2]_j,\Lambda]_{JM}$ and 
$[[j_2,j_1]_j,\Lambda]_{JM}$ sets of functions in the CR set, in short, labelled by $[[j_1,j_2]_j,\Lambda]_{JM}$.
To determine the irrep(s) corresponding to this representation, we calculated the characters for the group operations.

\emph{If the two monomers are in a different rotational state,} then any MS group operation that exchanges the two monomers has a zero character.
Furthermore, an operation $\hat{O} \in G_{576}$, which does not exchange the two monomer units, can be written as a product of operations acting on monomer `1' and `2', as $\hat{o}_1$ and $\hat{o}_2$, and an operation acting on the effective diatom, $\hat{d}$, hence $\hat{O}=\hat{o}_1\hat{o}_1\hat{d}$. 
Then, the character of any operation, $\hat{O}=\hat{o}_1\hat{o}_1\hat{d} \in G_{576}$, can be calculated for every $[[j_1,j_2]_j,\Lambda]_J$ representation from the character of the corresponding operation on the two monomers,  $\hat{o}_1, \hat{o}_2 \in T_{\rm d} (\rm M)$. 
Using these expressions, the characters for the any
$[[j_1,j_2]_j,\Lambda]_{JM}$ CR set of functions can be constructed. 

The symmetry assignment of the computed 6D intermolecular rovibrational states is carried out based on the CRD tables and the symmetry assignment of the CR functions.

\section{Raman transition moments \label{sec:raman}}

\subsection{Collection of the formulae}

The rovibrational absorption intensities can be expressed using the following working formula:\cite{ErRau20}
\begin{equation}
   I(\text{f}\leftarrow \text{i})= \frac{2\pi^2}{3} \frac{N_A}{\varepsilon_0 h^2 c^2}
   \frac{ e^{-E'/kT}(1-e^{-(E-E')/kT}) }{Q(T)} 
   (E-E') \mathcal{R} \; ,
\end{equation}
where $E$ and $E'$ are the energies of the initial, `i',  and final, `f', rovibrational states, respectively. 
Besides the well-known natural constants, $Q(T)$ denotes the partition function and $\mathcal{R}$ is the transition moment \cite{YuBaYa09} connecting the two rovibrational states, defined by 
\begin{equation}
\label{eq:transmoment}
    \mathcal{R} = \text{g}_{\text{ns}} 
    \sum_{M,M'} \sum_{A=X,Y,Z} 
    | \bra{\Psi^{\text{rv}}_{J'M'l'}} T_{A} 
    \ket{\Psi^{\text{rv}}_{JMl}} |^2.
\end{equation}
Using the same notation, $\ket{\Psi^{\text{rv}}_{JMl}}$ and $\ket{\Psi^{\text{rv}}_{J'M'l'}}$ correspond to the `i' and `f' rovibrational wave functions, respectively and $\text{g}_{\text{ns}}$ is the nuclear spin statistical weight factor. For simplicity, we define $S = \mathcal{R} / \text{g}_{\text{ns}}$. We also note that for an isolated molecular system, the rovibrational energy levels are degenerate with respect to the rotational quantum number, $M$, that describes the projection of the $\bos{J}$ angular momentum on the $Z$ axis of the laboratory (space-fixed) frame (LF). 

In Eq.~(\ref{eq:transmoment}), $T_{A}$ is the $A$ Cartesian component of the corresponding $T$ property tensor in the laboratory frame.
For instance, this tensor is the molecular dipole moment, $\mu_A$ $(A=X,Y,Z)$, for computing infrared intensities. 
Our rank-1 tensor implementation has been tested for the line strengths of the rovibrational transitions of the far-infrared and microwave spectrum of the \metwat\ dimer.\cite{DaAvMa21mw}
For Raman transitions, the property tensor has rank 2, \emph{i.e.}, a matrix $\alpha_{AB}$ $(A,B=X,Y,Z)$, and thus there are two Cartesian components ($A$ and $B$) in the space-fixed frame. 
The rovibrational integrals \cite{OwYa18} in Eq.~(\ref{eq:transmoment}) can be evaluated for a general $\Omega$-rank tensorial property according to
 \begin{equation}
 \label{eq:tensmatrix}
    \bra{\Psi^{\rm rv}_{J'M'l'}} T^{(\rm LF)}_A 
    \ket{\Psi^{\rm rv}_{JMl}} = \sum_{\omega=0}^\Omega
    \mathcal{M}_{A\omega}^{(J'M',JM)}
    \mathcal{K}_{\omega}^{(J'l',Jl)}
 \end{equation}
with
 \begin{equation}
  \begin{split}
    & \mathcal{M}_{A\omega}^{(J'M',JM)} = (-1)^{M'}
     \sqrt{(2J'+1)(2J+1)} \\
    & \times
     \sum_{\sigma=-\omega}^\omega 
     [U^{(\Omega)}]^{-1}_{A,\omega\sigma}
      \begin{pmatrix}
          J & \omega & J'\\
          M & \sigma & -M'
      \end{pmatrix}
    \end{split}
 \label{eq:tensmatrixM}
 \end{equation}
and
 \begin{equation}
 \begin{split}
    & \mathcal{K}_{\omega}^{(J'l',Jl)} = 
    \sum_{\substack{v,K,\tau\\v',K',\tau'}}
    [c_{v'K'\tau'}^{(J'l')}]^* c_{vK\tau}^{(Jl)}
     \sum_{\pm K', \pm K}[d_{K'}^{(\tau')}]^*
     d_K^{(\tau)}\\
     & \times (-1)^{K'}
     \sum_{\sigma=-\omega}^\omega \sum_{\alpha} 
      \begin{pmatrix}
          J & \omega & J'\\
          K & \sigma & -K'
      \end{pmatrix}
      U^{(\Omega)}_{\omega\sigma, \alpha}
     \bra{v'}T_\alpha^{(\rm BF)}\ket{v}.
 \end{split}
  \label{eq:tensmatrixK}
 \end{equation}
As a result, we need to numerically evaluate $\bra{v'}T_\alpha^{(\rm BF)}\ket{v}$-type integrals in the body-fixed (BF) frame using the vibrational `blocks' (corresponding to the different Wang functions) of the rovibrational 
wave function and the body-fixed expression of the $T_\alpha^\text{(BF)}$ property.

Within this approach, the BF integrals with respect to the internal coordinates are computed using the DVR grid. The property is evaluated at every grid point and then integrated with respect to the DVR vibrational basis.

The Raman intensities are calculated using the polarizability transitions with $\Omega=2$, and thus, there are two components:
the so-called isotropic (independent of the molecular orientation) with $\omega=0$ 
and the anisotropic with $\omega=2$. By using the general expressions, Eqs.~(\ref{eq:transmoment})--(\ref{eq:tensmatrixK}), the isotropic polarizability transition moment can be written as 
 \begin{equation}
 \begin{split}
    & \mathcal{R}_0 = \delta_{JJ'} \text{g}_{\text{ns}} 
    (2J'+1)(2J+1) \\
    & \times \Bigg| 
    \sum_{\substack{v,K,\tau\\v',K',\tau'}}
    [c_{v'K'\tau'}^{(J'l')}]^* c_{vK\tau}^{(Jl)}
     \sum_{\pm K', \pm K}[d_{K'}^{(\tau')}]^*
     d_K^{(\tau)}\\
     & \quad\quad \times (-1)^{K'}
     \sum_{\alpha\beta} 
      \begin{pmatrix}
          J & 0 & J'\\
          K & 0 & -K'
      \end{pmatrix}
      U^{(2)}_{00, \alpha\beta}
     \bra{v'}\alpha_{\alpha\beta}^{(\rm BF)}\ket{v}
    \Bigg |^2  \; ,
 \end{split}
 \label{eq:Riso}
 \end{equation}
where the summation over the $M$ and $M'$ quantum numbers was simplified according to
 \begin{equation}
 \label{eq:sum_m0}
    \sum_{M,M'}\sum_{A=X,Y,Z} \sum_{B=X,Y,Z} \Bigg|
     [U^{(2)}]^{-1}_{AB,00}
      \begin{pmatrix}
          J & 0 & J'\\
          M & 0 & -M'
      \end{pmatrix}
      \Bigg|^2 = \delta_{JJ'}.
 \end{equation}
 The 3$J$-symbols in Eqs.~(\ref{eq:Riso}) and (\ref{eq:sum_m0}) vanish unless $J=J'$. This leads to the selection rule $\Delta J=0$ for the isotropic transition moments, incorporated in the final equations by the Kronecker delta, $\delta_{JJ'}$.

For the anisotropic contribution ($\omega=2$), this summation over $M$ is always 1, and the final expression for the anisotropic transition moment is
 \begin{equation}
 \begin{split}
    & \mathcal{R}_2 = \text{g}_{\text{ns}} 
    (2J'+1)(2J+1) \\
    & \times \Bigg|
    \sum_{\substack{v,K,\tau\\v',K',\tau'}}
    [c_{v'K'\tau'}^{(J'l')}]^* c_{vK\tau}^{(Jl)}
     \sum_{\pm K', \pm K}[d_{K'}^{(\tau')}]^*
     d_K^{(\tau)}\\
     & \quad\quad \times (-1)^{K'}
     \sum_{\sigma=-2}^{2}
     \sum_{\alpha\beta} 
      \begin{pmatrix}
          J & 2 & J'\\
          K & \sigma & -K'
      \end{pmatrix}
     U^{(2)}_{2\sigma, \alpha\beta}
     \bra{v'}\alpha_{\alpha\beta}^{(\rm BF)}\ket{v}
    \Bigg |^2  .%\\
 \end{split}
 \label{eq:Raniso}
 \end{equation}
Again, as a direct consequence of the properties of the 3$J$-symbols, the anisotropic transitions are allowed if $\Delta J \leq \omega$, which introduces the selection rules, $\Delta J = 0, 1, 2$.
The non-zero matrix elements of $U^{(2)}_{\omega\sigma, \alpha\beta}$ and its `pseudo-'inverse $[U^{(2)}_{\alpha\beta,\omega\sigma}]^{-1}$ 
appearing in the equations for rank-2 properties 
are shown in the Supplementary Information.

\subsection{A simple polarizability model \label{sec:pol}}
We defined a simple polarizability model for \metmet\ to simulate its rovibrational Raman spectral features.

\begin{figure}%[h]
  \includegraphics[width=9cm]{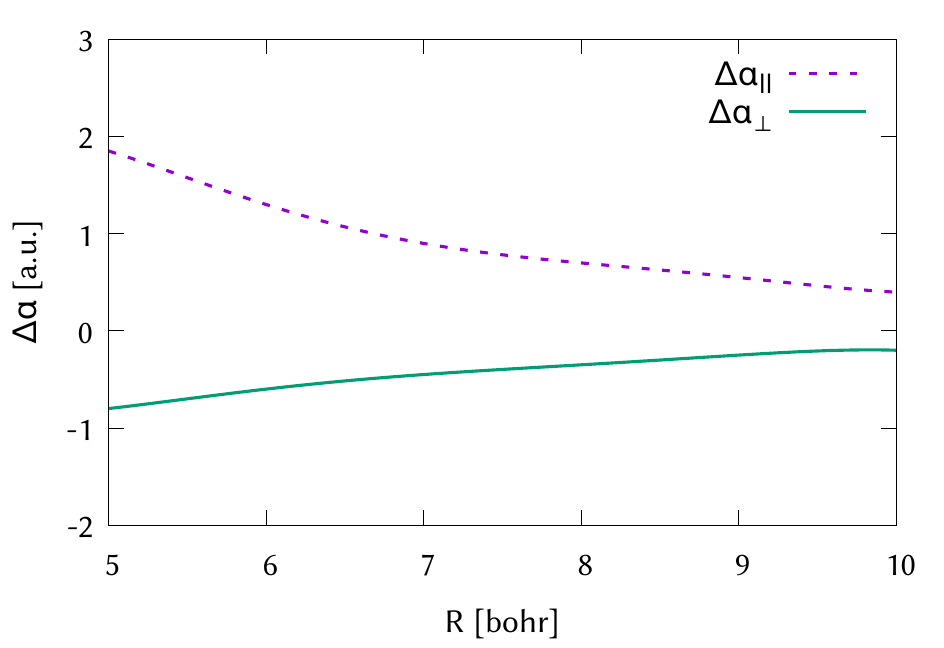}
  \caption{%
    Polarizability model constructed based on Ref.~\citenum{JeAsOs01}: 
    Radial dependence of both polarizability components of the methane dimer with respect to the $R$ intermolecular distance. The parallel polarizability ($\alpha_{\parallel}$) is represented with dash lines and perpendicular component ($\alpha_{\perp}$) with solid lines.
    \label{fig:alphaR}
    }
\end{figure}

In this model, we considered the methane molecules as spheres,
and the parallel ($\alpha_{\parallel}$)
and the perpendicular ($\alpha_{\perp}$)
polarizability components of the dimer were defined with respect to the axis connecting the two monomers, 
which was the $z$-axis for our particular choice of the body-fixed frame. 
Assuming negligible contribution of the interacting subsystems to the individual polarizabilities, the interaction polarizability of the dimer was defined as  
\begin{equation}
  \Delta \alpha = \alpha^{\text{\metmet}} - 2\alpha^{\text{CH$_4$}} \;,
  \label{eq:polapprox}
\end{equation}
where the value of the monomer polarizability is $\alpha^{\text{CH$_4$}} = 16.39~\text{a.u.}$ \cite{BuChKa10}.
Using this approximation, Eq.~(\ref{eq:polapprox}), Jensen \textit{et al.} \cite{JeAsOs01} computed both components of the interaction polarizability, $\Delta \alpha_{\perp}$ and $\Delta \alpha_{\parallel}$,  as a function of the intermolecular distance, $R$ (Fig.~\ref{fig:alphaR}). We have checked this simple model against explicit computations using the Dalton program package \cite{Dalton} for randomly positioned methane units, which confirmed the validity of this simple model. Otherwise, a polarizability surface can be developed, for example, along the lines of Ref.~\citenum{AvDaMa23}.

Since we consider in this work transitions between the lowest-energy states, the $R$ dependence of the interaction polarizability had a very small effect on the computed transition polarizabilities, and for this reason, we decided to use the values corresponding to the  equilibrium $R$ distance
$\Delta \alpha_{\parallel} = 1.3~\text{a.u.}$ 
and 
$\Delta \alpha_{\perp} = -0.6~\text{a.u.}$.
Within this model, the parallel and perpendicular components depend on the orientation of the body-fixed frame,
and for our body-fixed frame choice (Sec.~\ref{sec:coord}), the body-fixed polarizability tensor is written as
\begin{equation}
\alpha_{\alpha\beta}^{(\rm BF)} = 
    \begin{pmatrix}
          2\alpha^{\text{CH$_4$}} + \Delta \alpha_{\perp} & 0 & 0\\
          0 & 2\alpha^{\text{CH$_4$}} + \Delta \alpha_{\perp} & 0 \\
          0 & 0 & 2\alpha^{\text{CH$_4$}} + \Delta \alpha_{\parallel}
    \end{pmatrix} \; .
\end{equation}
In the present model, the body-fixed polarizability matrix (in atomic units) is a constant matrix over the intermolecular grid points:
\begin{equation}
\alpha_{\alpha\beta}^{(\rm BF)} = 
    \begin{pmatrix}
    32.18&0&0\\
    0&32.18&0\\
    0&0&34.08\\
    \end{pmatrix}\; . 
    \label{eq:pol_tensor}
\end{equation}

\clearpage

\section{Numerical results and discussion\label{sec:predlist}}
\begin{figure}
  \includegraphics[width=9cm]{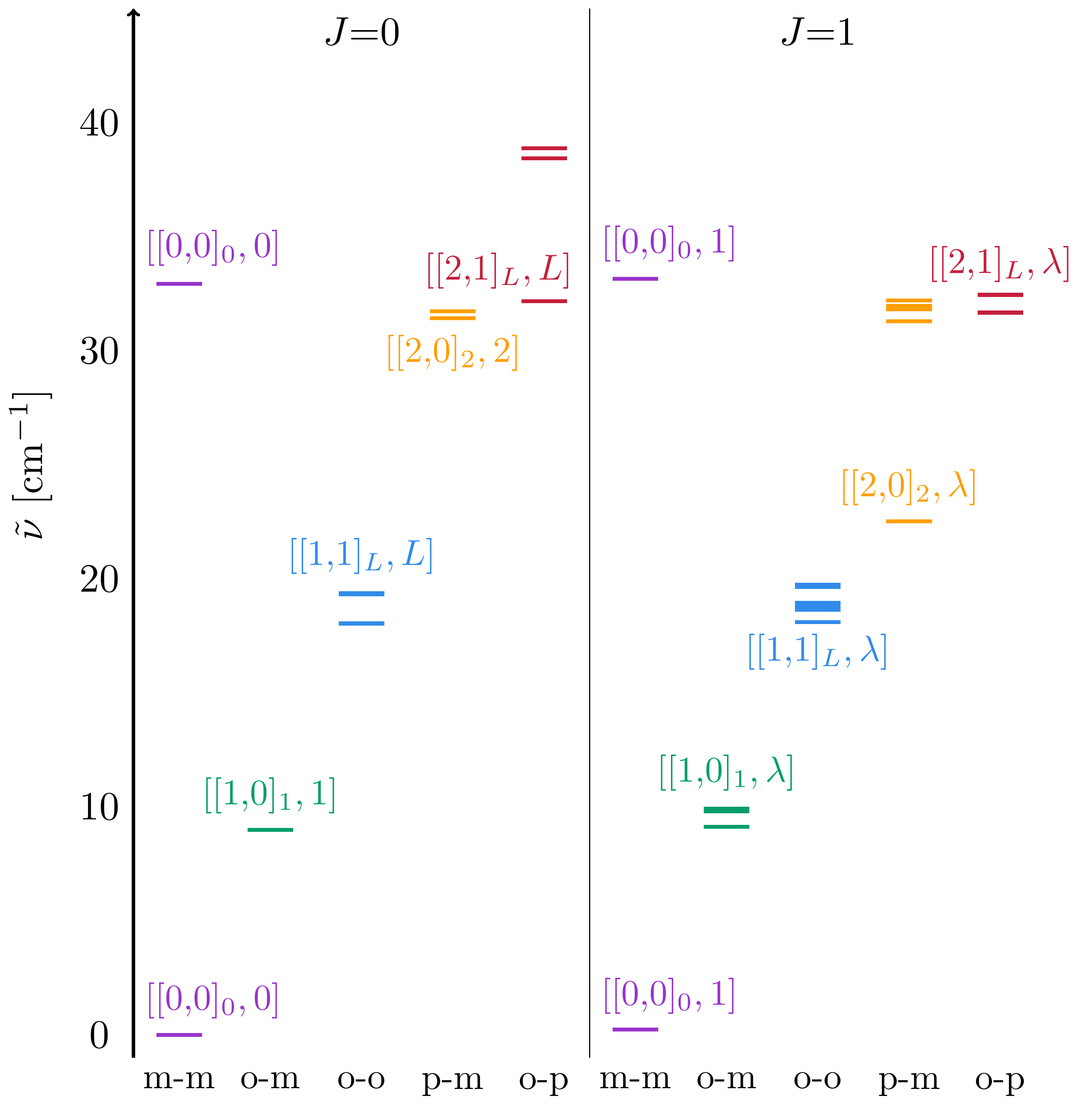} 
  \caption{%
    Energy level structure of the five lowest-energy spin isomers of the methane dimer with total angular quantum numbers $J$=0 and $J$=1. $[[j_1,j_2]_L,\Lambda]$ label the $j_1,j_2$ monomer and $\Lambda$ end-over-end rotational quantum numbers. 
    \label{fig:Jlevels}
  }
\end{figure}

\begin{figure}
  \includegraphics[width=9cm]{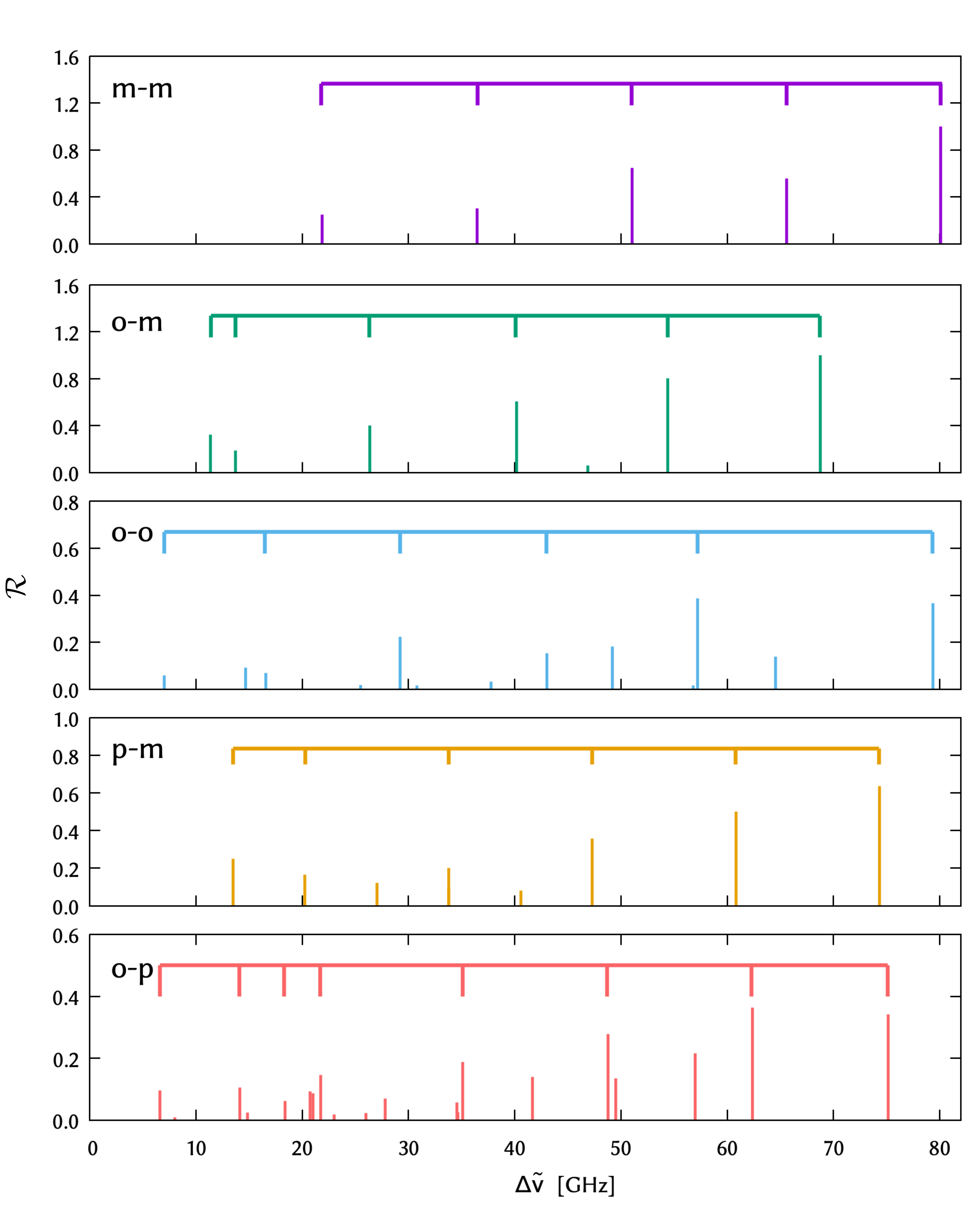}
  \caption{%
    Raman stick spectrum based on variational rovibrational computations. 
    $\mathcal{R}$ values were computed according to Eq.~(\ref{eq:transmoment}) and are plotted in atomic units. 
    \label{fig:Ramansticks}
  }
\end{figure}

Figure~\ref{fig:Jlevels} provides an overview of the $J=0$ and $J=1$ level structure for the five lowest-energy spin isomers. 
The idealized Raman `stick spectrum' (Fig.~\ref{fig:Ramansticks}) highlights potentially observable progressions from initial states corresponding to within ca. 1~\cm\ of the rovibrational ground state of the particular `spin isomer'.
For future potential experiments, the computed energy list and transition moments (up to $J=6$) are provided in the Supplementary Information, and they can then be used to simulate (even time-dependent) experimental conditions.

The main features of the energy level structure and a short discussion of the non-negligible transition moments are in order.

The meta-meta ground state is the absolute rovibrational ground state of this system and corresponds to the $j_1=j_2=0$ methane rotational quantum numbers and the $\Lambda=0$ end-over-end rotation quantum number. Correspondingly, the rotational Raman stick spectrum is predicted to feature a simple, regular progression, with observable transition moments with $\Delta J=2$ transitions (Table~\ref{tab:trans_m-m}).

In the ortho-meta ground state, one of the methane moieties is rotating `with a single quantum', $j_1=1,j_2=0$ (or $j_2=1,j_1=0$), which is coupled with the $\Lambda=1$ end-over-end rotation to obtain a $J=0$ `pure vibrational' state of the complex. Table~\ref{tab:trans_o-m} and Fig.~\ref{fig:Ramansticks} highlight a progression of regular $\Delta J=2$ transitions. An additional $\Delta J=1$ transition is also shown, which starts from a state that is very close in energy (less than 1~\cm\ separation) to the ground state of this spin isomer but with no end-over-end rotation, and hence a total $J=1$.

The ortho-ortho ground state is characterized by both methane fragments rotating with a single quantum, $j_1=j_2=1$, coupled (with $\Lambda=0$ or 2 end-over-end rotation) to $J=0$. Different coupling schemes can result in several close-lying energy levels with $J=1$ and $2$ (and higher) total rotational angular momentum quantum numbers. The numerous coupling possibilities of the rotors, end-over-end (de)excitation and overall rotational excitation give rise to multiple possible transition, and our computation predicts a longer list of non-negligible polarizability transition moments (Table~\ref{tab:trans_o-o}). Due to the various angular-momentum coupling options, we also find $\Delta J=0$ transition moments in the predicted list with non-negligible intensity.

The para-meta spin species is the next in the energetic ordering of the ground states with one methane `at rest' and the other `rotating with two quanta' ($j_1=0,j_2=2$ or $j_1=2,j_2=0$). Interestingly, the para-meta rovibrational ground state has $J=1$ total angular momentum quantum number. The $J=0$ vibrational ground state of this spin species is by $8.91$~\cm\ higher in energy than the $J=1$ p-m ground state and also higher in energy than the $J=2,3,4,5,6$ lowest vibrational energies of this spin isomer ($J>6$ states were not computed in this study).
Furthermore, it is interesting to add that the rigid-rotor analysis (Sec.~\ref{sec:klabels}) suggests that the $J=1$ absolute p-m ground state has $K=1$ label, whereas the next state in this block has $K=0$. So, the usual prolate symmetric top energy ordering is reversed and is reminiscent of `effective' oblate symmetric top features. All in all, the strongly fluxional character of the system and the very strong rovibrational couplings limit a simple rigid rotor analysis.
The significant polarizability transition moments following the regular $\Delta J=2$ progression from the $J=1$ ground state are collected in Table~\ref{tab:trans_p-m}, and a non-negligible $J=1$ to $J=2$ ($\Delta J=1$) transition is also predicted in the lowest-energy range of the spectrum.

Finally, the para-ortho species is the fifth in the energetic order, with one methane rotating with one quantum and the other with two quanta ($j_1=2,j_2=1$ or $j_1=1,j_2=2$) in the rovibrational ground state of this spin isomer. Similarly to the para-meta case, the para-ortho ground state is also a $J=1$ state, which is $-0.5$~\cm\ lower in energy than the lowest energy $J=0$ state of this spin isomer. 
The $J=1$ para-ortho ground state can be assigned to $K=1$, similarly to the p-m ground state, indicating a prolate to oblate transition of its `effective' properties. 
There are many possible coupling schemes of the two methane rotors and the end-over-end diatom angular momenta, which give rise to many non-negligible polarizability transition moments, including $\Delta J=2,1,$ and 0 cases (Table~\ref{tab:trans_p-o}, and a complicated stick spectrum in Fig.~\ref{fig:Ramansticks}). 

The highest-energy, para-para spin species was not identified among the computed  400 lowest-energy states with $J=0,1,2,3,4,5,6$.

All transition energies and polarizability moments computed in this work are deposited as Supplementary Information for potential use in conjunction with future experiments.

\begin{table}%[h]
\caption{%
Theoretically predicted Raman transitions 
for the meta-meta, $[[0,0]_j,\Lambda]_J$, spin isomer of \metmet.
$\mathcal{R}=g_\text{ns} S$ , Eqs.~(\ref{eq:Riso})--(\ref{eq:Raniso}), and the $g_\text{ns}$ nuclear spin statistical weight is 
15 for every even $J$ state and 10 for every odd $J$ state.
  \label{tab:trans_m-m}  
}
%\small{
\begin{tabular}{@{} l@{\ \ } l@{\ \ } l@{\ \ \ }  c@{\ \ \ \ } c@{\ \ \ \ } c@{\ \ } l@{\ \ }c@{\ \ } c@{}}
\hline\\[-0.35cm]
\hline\\[-0.25cm]
$(\text{J}.n)^{\rm f}$ & $\xleftarrow{}$ & $(\text{J}.n)^{\rm i}$ &
$\Delta\tilde\nu$~[\cm] &  $\Delta\tilde\nu$~[GHz]
& $S(\rm f \xleftarrow{} \rm i)$ \\ %& Comments$^\text{a}$ \\
\\[-0.25cm]
\hline\\[-0.25cm]
 J2.1 &  $\xleftarrow{}$ &   J0.1     &      \mynum{4}{0.7300924539}     &       \mynum{4}{21.887624}      &     \mynum{2}{2.406531826564458}   \\
 J4.1 &  $\xleftarrow{}$ &   J2.1     &      \mynum{4}{1.7022094903}     &       \mynum{4}{51.030964}      &     \mynum{2}{6.188089179934000}   \\
 J6.1 &  $\xleftarrow{}$ &   J4.1     &      \mynum{4}{2.6710889981}     &       \mynum{4}{80.077245}      &     \mynum{2}{9.844290646994574}   \\
 J3.1 &  $\xleftarrow{}$ &   J1.1     &      \mynum{4}{1.2164385123}     &       \mynum{4}{36.467914}      &     \mynum{2}{4.331719520267059}   \\
 J5.1 &  $\xleftarrow{}$ &   J3.1     &      \mynum{4}{2.1871719584}     &       \mynum{4}{65.569775}      &     \mynum{2}{8.021454558927154}   \\
\hline\\[-0.35cm]
\hline\\[-0.35cm]
\end{tabular}

\begin{flushleft}
%
%$^\text{a}$~Obs.: observed in the experiment.
%
\end{flushleft}
%}
\end{table}

\begin{table}%[h]
\caption{%
Theoretically predicted Raman transitions 
for the ortho-meta, $[[1,0]_j,\Lambda]_J$, spin isomer of \metmet.
The ground state corresponds to J0.2--7.
$\mathcal{R}=g_\text{ns} S$ , Eqs.~(\ref{eq:Riso})--(\ref{eq:Raniso}), and the $g_\text{ns}$ nuclear spin statistical weight is 
15 for every state.
  \label{tab:trans_o-m}  
}
%\small{
\begin{tabular}{@{} l@{\ \ } l@{\ \ } l@{\ \ \ }  c@{\ \ \ \ } c@{\ \ \ \ } c@{\ \ } l@{\ \ }c@{\ \ } c@{}}
\hline\\[-0.35cm]
\hline\\[-0.25cm]
$(\text{J}.n)^{\rm f}$ & $\xleftarrow{}$ & $(\text{J}.n)^{\rm i}$ &
$\Delta\tilde\nu$~[\cm] &  $\Delta\tilde\nu$~[GHz]
& $S(\rm f \xleftarrow{} \rm i)$ \\ %& Comments$^\text{a}$ \\
\\[-0.25cm]
\hline\\[-0.25cm]
J2.2--7   &  $\xleftarrow{}$ &   J0.2--7    &    \mynum{4}{0.4577738734}    &   \mynum{4}{13.723717}      &     \mynum{2}{1.814819379825325} \\ 
 J4.2--7   &  $\xleftarrow{}$ &   J2.2--7    &    \mynum{4}{1.3401242449}    &   \mynum{4}{40.175920}      &     \mynum{2}{5.799407062646929} \\ 
 J6.2--7   &  $\xleftarrow{}$ &   J4.2--7    &    \mynum{4}{2.2927891451}    &   \mynum{4}{68.736099}      &     \mynum{2}{9.573459696501383} \\ 
 J2.14--19 &  $\xleftarrow{}$ &   J0.2--7    &    \mynum{4}{1.5630151063}    &   \mynum{4}{46.858021}      &     \mynum{2}{0.589485209813700} \\ 
 J3.2--7   &  $\xleftarrow{}$ &   J1.2--7    &    \mynum{4}{0.8785115505}    &   \mynum{4}{26.337117}      &     \mynum{2}{3.838033115775553} \\ 
 J5.2--7   &  $\xleftarrow{}$ &   J3.2--7    &    \mynum{4}{1.8147422558}    &   \mynum{4}{54.404612}      &     \mynum{2}{7.702836995956829} \\ 
 J2.8--13  &  $\xleftarrow{}$ &   J1.14--19  &    \mynum{4}{0.3792668562}    &   \mynum{4}{11.370136}      &     \mynum{2}{3.106553101569780} \\
\hline\\[-0.35cm]
\hline\\[-0.35cm] 
\end{tabular}

\begin{flushleft}
%
%$^\text{a}$~Obs.: observed in the experiment.\\
%$^\text{b}$~Not observed perhaps due to other overlapping peaks or too weak transition. \\
%$^\text{c}$~Not observed due to too small initial population of states J1.14--19. \\
%
\end{flushleft}
%}
\end{table}

\begin{table}%[h]
\caption{%
Theoretically predicted Raman transitions 
for the ortho-ortho, $[[1,1]_j,\Lambda]_J$, spin isomer of \metmet.
The ground state corresponds to J0.8--16.
$\mathcal{R}=g_\text{ns} S$ , Eqs.~(\ref{eq:Riso})--(\ref{eq:Raniso}), and the $g_\text{ns}$ nuclear spin statistical weight is 
6 for every even $J$ state and 3 for every odd $J$ state.
  \label{tab:trans_o-o}  
}
%\small{
\begin{tabular}{@{} l@{\ \ } l@{\ \ } l@{\ \ \ }  c@{\ \ \ \ } c@{\ \ \ \ } c@{\ \ } l@{\ \ }c@{\ \ } c@{}}
\hline\\[-0.35cm]
\hline\\[-0.25cm]
$(\text{J}.n)^{\rm f}$ & $\xleftarrow{}$ & $(\text{J}.n)^{\rm i}$ &
$\Delta\tilde\nu$~[\cm] &  $\Delta\tilde\nu$~[GHz]
& $S(\rm f \xleftarrow{} \rm i)$ \\ %& Comments$^\text{a}$ \\
\\[-0.25cm]
\hline\\[-0.25cm]
J2.20--28  &  $\xleftarrow{}$ &   J0.8--16     &   \mynum{4}{0.2333059479}   &   \mynum{4}{ 6.994337}     &   \mynum{2}{1.413148422222392}   \\
 J4.20--28  &  $\xleftarrow{}$ &   J2.20--28    &   \mynum{4}{0.9741173192}   &   \mynum{4}{29.203307}     &   \mynum{2}{5.343773692771757}  \\
 J6.20--28  &  $\xleftarrow{}$ &   J4.20--28    &   \mynum{4}{1.9076803244}   &   \mynum{4}{57.190825}     &   \mynum{2}{9.261953426380055}  \\
 J2.56--64  &  $\xleftarrow{}$ &   J0.8--16     &   \mynum{4}{1.2602055064}   &   \mynum{4}{37.780016}     &   \mynum{2}{0.793296049518747}  \\
 J4.56--64  &  $\xleftarrow{}$ &   J2.56--64    &   \mynum{4}{1.6405587207}   &   \mynum{4}{49.182720}     &   \mynum{2}{4.359075809333542}  \\
 J6.56--64  &  $\xleftarrow{}$ &   J4.56--64    &   \mynum{4}{2.6473647811}   &   \mynum{4}{79.366011}     &   \mynum{2}{8.779941019942651}  \\
 J2.56--64  &  $\xleftarrow{}$ &   J2.20--28    &   \mynum{4}{1.0268995585}   &   \mynum{4}{30.785679}     &   \mynum{2}{0.394823513327492}  \\
 J2.65--73  &  $\xleftarrow{}$ &   J2.56--64    &   \mynum{4}{0.4889296185}   &   \mynum{4}{14.657743}     &   \mynum{2}{2.186940900275125}  \\
 J3.20--28  &  $\xleftarrow{}$ &   J1.20--28    &   \mynum{4}{0.5522380563}   &   \mynum{4}{16.555683}     &   \mynum{2}{3.284118956765719}  \\
 J5.20--28  &  $\xleftarrow{}$ &   J3.20--28    &   \mynum{4}{1.4351969539}   &   \mynum{4}{43.026128}     &   \mynum{2}{7.332532924410140}  \\
 J3.56--64  &  $\xleftarrow{}$ &   J1.20--28    &   \mynum{4}{1.8941845744}   &   \mynum{4}{56.786233}     &   \mynum{2}{0.724783613925550}  \\
 J5.56--64  &  $\xleftarrow{}$ &   J3.56--64    &   \mynum{4}{2.1526057950}   &   \mynum{4}{64.533507}     &   \mynum{2}{6.667383902906035}  \\
 J1.56--64  &  $\xleftarrow{}$ &   J1.20--28    &   \mynum{4}{0.8503863158}   &   \mynum{4}{25.493944}     &   \mynum{2}{0.858461614018417}  \\
\hline\\[-0.35cm]
\hline\\[-0.35cm] 
\end{tabular}

\begin{flushleft}
\end{flushleft}
%}
\end{table}

\begin{table}[h]
\caption{%
Theoretically predicted Raman transitions 
for the para-meta, $[[2,0]_j,\Lambda]_J$, spin isomers of the \metmet.
The ground state corresponds to J1.83--86.
$\mathcal{R}=g_\text{ns} S$ , Eqs.~(\ref{eq:Riso})--(\ref{eq:Raniso}), and the $g_\text{ns}$ nuclear spin statistical weight is 
10 for every state.
  \label{tab:trans_p-m}  
}
%\small{
\begin{tabular}{@{} l@{\ \ } l@{\ \ } l@{\ \ \ }  c@{\ \ \ \ } c@{\ \ \ \ } c@{\ \ } l@{\ \ }c@{\ \ } c@{}}
\hline\\[-0.35cm]
\hline\\[-0.25cm]
$(\text{J}.n)^{\rm f}$ & $\xleftarrow{}$ & $(\text{J}.n)^{\rm i}$ &
$\Delta\tilde\nu$~[\cm] &  $\Delta\tilde\nu$~[GHz]
& $S(\rm f \xleftarrow{} \rm i)$ \\ %& Comments$^\text{a}$ \\
\\[-0.25cm]
\hline\\[-0.25cm]
 J3.101--104  &  $\xleftarrow{}$ &   J1.83--86    &   \mynum{4}{1.1264576870}     &   \mynum{4}{33.770357}     &  \mynum{2}{2.873299920156180}  \\
 J5.101--104  &  $\xleftarrow{}$ &   J3.101--104   &   \mynum{4}{2.0286150792}     &   \mynum{4}{60.816359}     &  \mynum{2}{7.185101450668927} \\
 J4.101--104  &  $\xleftarrow{}$ &   J2.101--104   &   \mynum{4}{1.5773954411}     &   \mynum{4}{47.289132}     &  \mynum{2}{5.131491033679453} \\
 J6.101--104  &  $\xleftarrow{}$ &   J4.101--104   &   \mynum{4}{2.4801032396}     &   \mynum{4}{74.351635}     &  \mynum{2}{9.146104248002114} \\
 J2.101--104  &  $\xleftarrow{}$ &   J1.83--86    &   \mynum{4}{0.4505287845}     &   \mynum{4}{13.506515}     &  \mynum{2}{3.577553125733367}  \\
 J3.101--104  &  $\xleftarrow{}$ &   J2.101--104   &   \mynum{4}{0.6759289025}     &   \mynum{4}{20.263842}     &  \mynum{2}{2.370103371078527} \\
 J4.101--104  &  $\xleftarrow{}$ &   J3.101--104   &   \mynum{4}{0.9014665386}     &   \mynum{4}{27.025291}     &  \mynum{2}{1.762185462391667} \\
 J5.101--104  &  $\xleftarrow{}$ &   J4.101--104   &   \mynum{4}{1.1271485406}     &   \mynum{4}{33.791068}     &  \mynum{2}{1.394265831166427} \\
 J6.101--104  &  $\xleftarrow{}$ &   J5.101--104   &   \mynum{4}{1.3529546990}     &   \mynum{4}{40.560567}     &  \mynum{2}{1.146501746630578} \\
\hline\\[-0.35cm]
\hline\\[-0.35cm] 
\end{tabular}

\begin{flushleft}
%
%$^\text{a}$~
%
\end{flushleft}
%}
\end{table}

\begin{table}[h]
\caption{%
Theoretically predicted Raman transitions 
for the para-ortho, $[[2,1]_j,\Lambda]_J$, spin isomer of \metmet.
The ground state corresponds to J1.93--104.
$\mathcal{R}=g_\text{ns} S$ , Eqs.~(\ref{eq:Riso})--(\ref{eq:Raniso}), and the $g_\text{ns}$ nuclear spin statistical weight is 
6 for every state.
%\madd{$\mathcal{R}=g_\text{ns} S$, Eqs.~(\ref{eq:Riso})--(\ref{eq:Raniso}), and the $g_\text{ns}$ nuclear spin statistical weight is ... } {\color{green} relabel $S_0$ to S in the table?}
  \label{tab:trans_p-o}  
}
%\small{
\begin{tabular}{@{} l@{\ \ } l@{\ \ } l@{\ \ \ }  c@{\ \ \ \ } c@{\ \ \ \ } c@{\ \ } l@{\ \ }c@{\ \ } c@{}}
\hline\\[-0.35cm]
\hline\\[-0.25cm]
$(\text{J}.n)^{\rm f}$ & $\xleftarrow{}$ & $(\text{J}.n)^{\rm i}$ &
$\Delta\tilde\nu$~[\cm] &  $\Delta\tilde\nu$~[GHz]
& $S(\rm f \xleftarrow{} \rm i)$ \\ %& Comments$^\text{a}$ \\
\\[-0.25cm]
\hline\\[-0.25cm]
 J0.45--56      & $\xleftarrow{}$ &  J2.127--138  &  \mynum{4}{0.2669660631}      &   \mynum{4}{8.003442}     &  \mynum{2}{0.221240670837358}     \\
 J4.127--138  & $\xleftarrow{}$ &   J2.127--138   &  \mynum{4}{1.1709132328 }      &   \mynum{4}{35.103101}     &  \mynum{2}{4.491676153746305}   \\
 J6.121--132  & $\xleftarrow{}$ &   J4.127--138   &  \mynum{4}{2.0801697323 }      &   \mynum{4}{62.361928}     &  \mynum{2}{8.711321511160484}   \\
 J2.149--160  & $\xleftarrow{}$ &   J2.127--138   &  \mynum{4}{0.6136907711 }      &   \mynum{4}{18.397989}     &  \mynum{2}{1.479547767064127}   \\
 J2.161--172  & $\xleftarrow{}$ &   J2.149--160   &  \mynum{4}{0.4951957358 }      &   \mynum{4}{14.845597}     &  \mynum{2}{0.573870422668730}   \\
 J4.143--154  & $\xleftarrow{}$ &   J2.149--160   &  \mynum{4}{1.6511445498 }      &   \mynum{4}{49.500075}     &  \mynum{2}{3.210109213430225}   \\
 J6.139--150  & $\xleftarrow{}$ &   J4.143--154   &  \mynum{4}{2.5059897096 }      &   \mynum{4}{75.127692}     &  \mynum{2}{8.195522396166105}   \\
 J4.143--154  & $\xleftarrow{}$ &   J2.161--172   &  \mynum{4}{1.1559488140 }      &   \mynum{4}{34.654478}     &  \mynum{2}{0.608076877494687}   \\
 J4.162--173  & $\xleftarrow{}$ &   J2.161--172   &  \mynum{4}{1.9003683442 }      &   \mynum{4}{56.971618}     &  \mynum{2}{5.167683671986127}   \\
 J6.162--173  & $\xleftarrow{}$ &   J4.162--173   &  \mynum{4}{2.8413881355 }      &   \mynum{4}{85.182685}     &  \mynum{2}{9.043298370712638}   \\
 J3.127--138  & $\xleftarrow{}$ &   J1.93--104    &  \mynum{4}{0.6913771635 }      &   \mynum{4}{20.726969}     &  \mynum{2}{2.207067553745063}   \\
 J5.127--138  & $\xleftarrow{}$ &   J3.127--138   &  \mynum{4}{1.6273200312 }      &   \mynum{4}{48.785834}     &  \mynum{2}{6.658627769141455}   \\
 J1.121--132  & $\xleftarrow{}$ &   J1.93--104    &  \mynum{4}{0.7673105689 }      &   \mynum{4}{23.003395}     &  \mynum{2}{0.426597146708457}   \\
 J3.162--173  & $\xleftarrow{}$ &   J1.121--132   &  \mynum{4}{1.3902789546 }      &   \mynum{4}{41.679520}     &  \mynum{2}{3.333309112719472}   \\
 J3.149--160  & $\xleftarrow{}$ &   J3.127--138   &  \mynum{4}{0.8677513309 }      &   \mynum{4}{26.014534}     &  \mynum{2}{0.536708349357072}   \\
 J2.127--138  & $\xleftarrow{}$ &   J1.93--104    &  \mynum{4}{0.2207617511 }      &   \mynum{4}{ 6.618272}     &  \mynum{2}{2.280753780167315}   \\
 J3.127--138  & $\xleftarrow{}$ &   J2.127--138   &  \mynum{4}{0.4706154124 }      &   \mynum{4}{14.108697}     &  \mynum{2}{2.521758889548672}   \\
 J4.127--138  & $\xleftarrow{}$ &   J3.127--138   &  \mynum{4}{0.7002978204 }      &   \mynum{4}{20.994403}     &  \mynum{2}{2.055500383566622}   \\
 J5.127--138  & $\xleftarrow{}$ &   J4.127--138   &  \mynum{4}{0.9270222108 }      &   \mynum{4}{27.791431}     &  \mynum{2}{1.664349499489075}   \\
 J6.121--132  & $\xleftarrow{}$ &   J5.127--138   &  \mynum{4}{1.1531475214 }      &   \mynum{4}{34.570498}     &  \mynum{2}{1.370381992772167}   \\
 J3.149--160  & $\xleftarrow{}$ &   J2.149--160   &  \mynum{4}{0.7246759722 }      &   \mynum{4}{21.725242}     &  \mynum{2}{3.486978888442655}   \\
\hline\\[-0.35cm]
\hline\\[-0.35cm] 
\end{tabular}
\end{table}

\section{Summary and conclusion}
%-----------------------------------
This paper reported rovibrational computations for the methane dimer on an \emph{ab initio} intermolecular potential energy surface.

The equilibrium structure of \metmet\ is a (prolate) symmetric top. $K$ labels can be unambiguously assigned to the lowest-energy states up to $J=2-3$. The lowest-energy rotational states of the meta-meta ($I=0$), ortho-meta ($I=1$ and 0), and ortho-ortho ($I=1$) proton spin isomers corresponding to $[j_1,j_2]=[0,0]$, $[0,1]$, and $[1,1]$ rotational quanta assignable to the methane subunits show prolate-type energetic ordering. This ordering is apparently reversed for the [2,0] para-meta ($I=2$ and 0) and [2,1] para-ortho ($I=2$ and 1) spin `isomers', rendering an oblate-type energetic ordering for the lowest-energy rotational states. The strongly fluxional character of this weakly bound complex limits the applicability of rigid-rotor-type concepts and the weakly coupled rotor picture can be used more naturally.

To facilitate future detection and assignment of the intermolecular rotational transitions of this simplest representative of alkyl-alkyl interactions, potentially by some Raman-type spectroscopic technique, a simple polarizability model was designed and used to predict polarizability transition moments. Simple rotational Raman progressions are predicted for the meta-meta and ortho-meta species, while more features in the para-meta, and a more complicated pattern can be expected for the ortho-ortho and para-ortho species.

\section*{Acknowledgements}
We thank the financial support of the Hungarian National Research, Development, and Innovation Office (FK 142869).
%%%END OF MAIN TEXT%%%